\documentclass[journal,two columns, 12px]{IEEEtran}
\ifCLASSINFOpdf
  
\else

\fi
\usepackage{cellspace}
\usepackage{cite}
\usepackage{amsmath,amssymb,amsfonts}
\usepackage{algorithmic}
\usepackage{graphicx,balance}
\usepackage{subfig}
\usepackage{balance}
\usepackage{textcomp}
\usepackage{xcolor, soul}
\sethlcolor{green}
\usepackage{color}
\usepackage{url}
\usepackage{xspace}
\usepackage{booktabs}
\usepackage{comment}
\usepackage{acronym}
\usepackage[none]{hyphenat}
\usepackage{ragged2e}
\usepackage{tabularx}
\usepackage{tikz}
\usetikzlibrary{mindmap,trees,arrows}
\newcommand{\tabitem}{~~\llap{\textbullet}~~}

\newcommand{\etc}{etc.\@\xspace}

\def\BibTeX{{\rm B\kern-.05em{\sc i\kern-.025em b}\kern-.08em
    T\kern-.1667em\lower.7ex\hbox{E}\kern-.125emX}}

\usepackage{hyperref}

\hypersetup{
    colorlinks=true,        
    linkcolor=red,         
    citecolor=blue,         
    urlcolor=blue           
}

\acrodef{HAT}{Human-Autonomy Teaming}
\acrodef{AI}{Artificial Intelligence}
\acrodef{XAI}{Explainable AI}
\acrodef{FL}{Federated Learning}
\acrodef{ML}{Machine Learning}
\acrodef{DL}{Deep Learning}
\acrodef{NLP}{Natural Language Processing}
\acrodef{MRI}{Magnetic Resonance Imaging}
\acrodef{RCO}{Reduced Crew Operations}
\acrodef{UAM}{Urban Air Mobility}
\acrodef{UAVs}{Unmanned Aerial Vehicles}
\acrodef{GANs}{Generative Adversarial Networks}
\acrodef{MDO}{Multi-Domain Operations}
\acrodef{RNNs}{Recurrent Neural Networks}
\acrodef{CNNs}{Convolutional Neural Networks}
\acrodef{k-NN}{k-Nearest Neighbors}
\acrodef{HMI}{Human-Machine Interface}
\acrodef{HMT}{Human-Machine Teaming}
\acrodef{ASIST}{Artificial Social Intelligence for Successful Teams}
\acrodef{GDPR}{General Data Protection Regulation}

\begin{document}

\title{AI-Driven Human-Autonomy Teaming in Tactical Operations: Proposed Framework, Challenges, and Future Directions

}

\author{\IEEEauthorblockN{Desta Haileselassie Hagos, Hassan El Alami, Danda B. Rawat}

\IEEEauthorblockA{Howard University, Washington, DC, USA \\ \{desta.hagos, hassan.elalami, danda.rawat\}@howard.edu} 
}

\renewcommand\IEEEkeywordsname{Keywords}

\markboth{Submitted for review to the Proceedings of the IEEE}
{Hagos \MakeLowercase{\textit{et al.}}: AI-Driven Human-Autonomy Teaming in Tactical Operations: Proposed Framework, Challenges, and Future Directions}

\maketitle

\begin{abstract}

\ac{AI} techniques, particularly machine learning techniques, are rapidly transforming tactical operations by augmenting human decision-making capabilities. This paper explores \ac{AI}-driven \ac{HAT} as a transformative approach, focusing on how it empowers human decision-making in complex environments. While trust and explainability continue to pose significant challenges, our exploration focuses on the potential of AI-driven \ac{HAT} to transform tactical operations. By improving situational awareness and supporting more informed decision-making, AI-driven HAT can enhance the effectiveness and safety of such operations. To this end, we propose a comprehensive framework that addresses the key components of AI-driven \ac{HAT}, including trust and transparency, optimal function allocation between humans and \ac{AI}, situational awareness, and ethical considerations. The proposed framework can serve as a foundation for future research and development in the field. By identifying and discussing critical research challenges and knowledge gaps in this framework, our work aims to guide the advancement of AI-driven \ac{HAT} for optimizing tactical operations. We emphasize the importance of developing scalable and ethical AI-driven HAT systems that ensure seamless human-machine collaboration, prioritize ethical considerations, enhance model transparency through \ac{XAI} techniques, and effectively manage the cognitive load of human operators.

\end{abstract}

\begin{IEEEkeywords}
Artificial Intelligence, Human-Autonomy Teaming, Human-Machine Teaming, Tactical Operations
\end{IEEEkeywords}

\section{Introduction}
\label{introduction}

\IEEEPARstart{T}{he} convergence of \ac{AI} and autonomous technologies has revolutionized various industries, including defense and tactical operations. The rise of \ac{HAT} can be attributed to several factors, including rapid advancements in autonomous technologies and \ac{AI}~\cite{o2023human}, the increasing complexity of tasks and environments, the development of more capable autonomous systems, and the increasing availability of data and computing power~\cite{o2022human}. As these technologies have become more sophisticated and capable, there has been a growing recognition of the potential collaborations that can be achieved by combining human cognitive abilities with the computational power and efficiency of autonomous systems~\cite{lyons2021human}. The rise of modern \ac{HAT} systems has also been driven by the need to address the complexities and challenges of rapidly evolving and dynamic environments. As tasks become more complex, time-sensitive, and data-intensive, the collaboration between humans and autonomous agents becomes crucial for effectively navigating and responding to these challenges.

\ac{HAT} is an emerging field that explores collaborative partnerships between humans and autonomous systems to perform tasks or achieve common goals~\cite{o2022human, johnson2012autonomy, wynne2018integrative, chen2016human}. This involves a collaborative arrangement in which at least one human worker collaborates with one or more autonomous agents~\cite{o2022human}. This collaborative approach has the potential to revolutionize how tasks are accomplished across various sectors and pave the way for a future where humans and intelligent autonomous systems will work hand in hand to tackle complex problems and achieve shared goals. \ac{HAT} systems are designed to allow humans to delegate tasks to intelligent autonomous agents while maintaining overall mission control~\cite{endsley2017here}. Autonomous agents, in this context, refer to computer entities with varying degrees of self-governance in decision-making, adaptation, and communication. This definition has been supported by studies conducted by the research works in~\cite{mercado2016intelligent,myers2018autonomous}. The integration of human cognitive capabilities with the computational power and efficiency of autonomous systems in \ac{HAT} enhances performance, decision-making, and overall system capabilities.

Here, we define and clarify some key concepts that are fundamental to understanding the scope and context of this study. These concepts include \ac{AI}, Autonomy, Autonomous Systems, and Tactical Autonomy. By providing clear definitions and distinguishing between these terms, we aim to establish a common understanding among our readers.

\vspace{1.0ex}

\noindent \textbf{Autonomy}. Autonomy in the context of \ac{HAT} describes the ability of intelligent autonomous systems or agents to operate and make decisions independently in a team setting with varying degrees of self-governance~\cite{lyons2021human, klare2019autonomous}. This involves a higher degree of decision-making capability in autonomous systems based on learning, adaptation, and reasoning. It is a property of a system, not a technology itself~\cite{klare2019autonomous}. An autonomous entity can perceive, reason, plan, and act in pursuit of specific goals or objectives without constant human intervention. It is important to note that the level of autonomy can vary, ranging from fully autonomous systems that make all their decisions to semi-autonomous systems that require human input at certain points~\cite{klare2019autonomous}. In the context of tactical autonomy, \ac{HAT} involves the integration of autonomous capabilities into tactical operations. This integration can include various applications, such as using autonomous systems to gather intelligence, perform surveillance, and perform other critical activities. Autonomy enables systems to operate in complex and uncertain environments, learn from experience, and make decisions without explicit human intervention in every scenario. However, it is important to distinguish this from traditional automation, which typically follows pre-programmed rules, decision trees, or logic-based algorithms to perform tasks or make decisions. Traditional automation has limited adaptability and flexibility to handle dynamic or unforeseen situations without explicit programming. This paper discusses how AI-driven autonomy differs from traditional automation by emphasizing learning, adaptation, and decision-making capabilities. These capabilities ultimately enhance the overall effectiveness and agility of human-autonomy teaming in tactical operations.

\vspace{1.0ex}

\noindent \textbf{Autonomous Systems}. Autonomous systems can perform tasks or operations without constant human control. They utilize \ac{AI} algorithms and sensors to perceive and navigate their environment, achieving a high degree of autonomy~\cite{watson2005autonomous}.

\vspace{1.0ex}
\noindent \textbf{Tactical Autonomy}. In this study, tactical autonomy refers to autonomous systems' ability to make real-time decisions and take actions in dynamic and complex operational environments~\cite{hagos2022recent}. This involves the seamless coordination and interaction between humans and autonomous systems, enabling them to function as a unified team with complementary strengths~\cite{hagos2022recent}. \ac{HAT} focuses on achieving shared mission goals through seamless coordination and collaboration between human operators and intelligent autonomous systems~\cite{shively2022human}. This paper introduces an AI-driven \ac{HAT}, which integrates \ac{AI} into HAT frameworks. This approach improves decision-making, situational awareness, and operational effectiveness by combining the strengths of human expertise and AI capabilities. Tactical autonomy, which combines human cognitive abilities, such as adaptability, intuition, and creativity, with the computational power, precision, and dynamic execution of autonomous systems, has the potential to revolutionize various fields, including defense, emergency response, law enforcement, and hazardous environments~\cite{hagos2022recent}. It is important to differentiate between tactical and strategic autonomy to clarify how AI-driven human-autonomy teaming contributes to both levels of autonomy in military and operational contexts. Strategic autonomy refers to a nation or organization's ability to make autonomous choices regarding broad security goals, whereas tactical autonomy, in contrast to strategic autonomy, focuses on individual units or teams acting independently within a specific mission~\cite{beaucillon2023special}. Strategic autonomy involves higher-level decision-making and planning that considers long-term goals, overall mission objectives, and broader situational awareness. It addresses the coordination, allocation of resources, and strategic decision-making processes that guide the overall mission or campaign~\cite{beaucillon2023special}.

\vspace{1.0ex}
\noindent \textbf{Tactical Operations}. Tactical operations involve coordinated activities in a specific area or environment, typically in a military, law enforcement, or strategic context, focusing on achieving short-term objectives through rapid decision-making, adaptation to dynamic situations, and the application of military skills and resources within a localized area and timeframe~\cite{united1994department}.

In recent years, advancements in \ac{AI}, \ac{ML}, robotics, and sensor technologies have paved the way for realizing the potential of tactical autonomy~\cite{hagos2022recent}. These technological advancements have enabled autonomous systems to perform complex tasks, process vast amounts of data in real-time, make informed decisions, and collaborate with human team members seamlessly~\cite{hagos2022recent}. This has opened new possibilities for augmenting human capabilities, optimizing resource allocation, and improving overall operational efficiency. However, effective tactical autonomy requires a comprehensive understanding of the dynamics between humans and autonomous systems. Human factors, including trust, communication, shared situational awareness, and decision-making, play a vital role in ensuring successful \ac{HAT}. Challenges such as establishing appropriate levels of trust, addressing potential cognitive biases, managing workload distribution, and maintaining effective communication channels must be carefully addressed to ensure seamless collaboration and maximize the potential benefits of tactical autonomy. \ac{HAT} for tactical autonomy is a collaborative approach to using humans and autonomous systems to operate and control weapons and other military systems. In \ac{HAT}, the human operators and autonomous systems work together to achieve common goals. The human operators are responsible for the overall mission and making high-level decisions. Autonomous systems are responsible for performing assigned tasks.

As explained in detail in Section~\ref{characteristics_key_components}, human operators contribute strategic insight, context, and high-level decision-making capabilities based on their experience and understanding of the mission's goals. The interaction and communication represent the interfaces and communication channels through which each component exchanges information, collaborates, and makes joint decisions. Within the context of a shared decision-making process, human operators and autonomous systems engage in a collaborative decision-making process, sharing insights, data, and recommendations to formulate effective strategies. The autonomous system is responsible for real-time data processing, analysis, and execution of specific tasks supporting human operators with timely and pertinent information. Subsequently, once decisions are made, the autonomous system performs specific tasks, including reconnaissance, navigation, or data collection, in alignment with the directives of the shared decision-making process.

This paper comprehensively explores the historical development and current state of \ac{HAT} and delves into the opportunities, challenges, and potential future directions in leveraging \ac{AI} for tactical autonomy. It emphasizes the transformative impact of \ac{AI} on tactical autonomy and presents opportunities for improved decision-making, situational awareness, and resource optimization. By acknowledging and addressing the challenges associated with \ac{AI} adoption, and by charting future directions for research, we can pave the way for a future where humans and autonomous systems seamlessly collaborate, ultimately leading to safer, more efficient, and successful missions in tactical environments.

\subsection{Scope and Contributions}

The main contribution of this paper is its forward-looking study of the applications, trends, and disruptive technologies that will drive the \ac{HAT} revolution in complex and dynamic environments. This provides a clear picture of \ac{HAT} services and practical recommendations for future work. 

\subsection{Contributions}

This paper makes the following key contributions to the field of \ac{HAT}.

\vspace{1.0ex}
\begin{itemize}
    \item We propose a comprehensive conceptual framework for AI-driven HAT in tactical operations, describing critical components such as trust and transparency, function allocation, situational awareness, and ethical considerations. The proposed framework provides a foundation to understand and advance the integration of AI into HAT for tactical environments.
    
    \item We provide a comprehensive overview of the opportunities and key challenges associated with incorporating AI-driven HAT into tactical operations.

    \item We explore the symbiotic relationship between AI and HAT, presenting a thorough analysis of how AI-driven HAT enhances decision-making, situational awareness, and operational effectiveness in tactical environments.

    \item We identify several research directions for future work in AI-driven HAT, emphasizing ethical considerations, building transparent AI models, and advancing human-centric design principles to fully realize the potential of tactical autonomy.

\end{itemize}

Table~\ref{table:comparison} compares our work to existing studies. In this paper, we explore and address research questions related to AI-driven HAT to enhance tactical operations, covering various aspects and challenges.

\begin{itemize}
    \item How do \ac{AI} and \ac{HAT} benefit each other when achieving tactical autonomy?   
    
    \item What are the main opportunities and challenges associated with incorporating AI-driven HAT in the context of tactical operations? 
    
    \item How can AI-driven \ac{HAT} be best used in tactical operations to improve success and decision-making?
    
    \item What is the plan for AI-driven HAT and how can it improve the collaboration between humans and autonomous systems in tactical situations?  
    
    \item How can AI-driven \ac{HAT} help humans and autonomous systems work together smoothly to achieve common goals in tactical environments?  
    
    \item What ethical concerns must be considered when developing and using AI-driven HAT systems? 
    
    \item How can we make AI models in HAT more understandable, and why does this matter for better decision-making and trust in autonomous systems?
    
    \item What design principles should be followed to create user-friendly AI-driven HAT systems for human operators in tactical settings?
\end{itemize}

\begin{table*}

\caption{Comparison of our work to existing works.}
\label{table:comparison}
\begin{tabularx}{\linewidth}{r|l| X} 
\midrule
\textbf{Year} & \textbf{Publications} & \textbf{Main Research Focus and Scope}\\\midrule

2018 & Ref~\cite{demir2018team} & \tabitem Explores the relationship between team coordination dynamics and team performance for human-autonomy teams using an extended version of nonlinear dynamical systems methods. \\ \midrule

2018 & Ref~\cite{shively2018human} & \tabitem Proposed a framework for \ac{HAT}, incorporating three key tenets: transparency, bi-directional communication, and operator-directed authority. \\ \midrule

2019 & Ref~\cite{roth2019function} & \tabitem Discusses what function allocation and challenges in allocating tasks between humans and autonomous machines. \\ \midrule

2020 & Ref~\cite{schelble2020towards} & \tabitem Provides a framework for practitioners to make informed decisions regarding the integration and training of human-autonomy teams in applied settings. \\ \midrule

2020 & Ref~\cite{lucero2020human} & \tabitem Proposes a new approach to using ML agents in real-time strategy games to collaborate with human players rather than competing against them. \\ \midrule

2021 & Ref~\cite{lyons2021human} & \tabitem Examines the differences between automation and autonomy and how insights from human-human teaming can be applied to \ac{HAT}. The authors have identified research gaps that need to be addressed to improve the understanding of HAT.\\ \midrule

2022 & Ref~\cite{o2022human} & \tabitem Provides a comprehensive understanding of the research environment, dependent variables, independent variables, key findings, and future research directions related to human-autonomy teamwork. \\ \midrule

2022 & Ref~\cite{teaming2022state} & \tabitem Emphasizes the need for humans and AI to work together effectively, particularly in complex situations. It  examines the factors affecting the design and implementation of AI systems for human interaction. In addition, it provides a detailed roadmap for future HAT research, particularly emphasizing the perspectives of human factors, which aligns well with our focus on enhancing tactical operations through AI-driven HAT.  \\ \midrule

2024 & Our Paper & \tabitem Proposes a comprehensive conceptual framework for AI-driven HAT in tactical operations, detailing critical components, such as trust and transparency, function allocation, situational awareness, and ethical considerations.

\tabitem Explores the advantages and challenges associated with integrating AI-powered \ac{HAT} into tactical operations.

\tabitem Provides a thorough exploration of the symbiotic relationship between \ac{AI} and \ac{HAT} in the context of tactical operations.

\tabitem Identifies several research directions, including ethical considerations, building transparent AI models, and advancing human-centric design principles, for future work in AI-driven HAT. 
\\ \midrule

\end{tabularx}

\end{table*}

\subsection{Methodology}
This study investigates the potential of AI-driven HAT to revolutionize tactical operations. To achieve this, we conducted a systematic literature review to identify and analyze relevant academic research. Our search primarily targeted prominent academic databases such as Google Scholar, IEEE Xplore, ACM Digital Library, and ScienceDirect for scholarly articles published up to 2024. We focused on studies published up to May 2024 that emphasized empirical research and theoretical frameworks to explore the application of AI in human-autonomy teaming for tactical operations. Note that studies that focused on general AI applications without a tactical operation context were excluded. We employed a combination of keywords, including ``AI-driven human-autonomy teaming,'' ``tactical operations,'' ``situational awareness,'' ``automated decision-making,'' ``Integrating AI and HAT,'' ``situation models,'' and ``shared situational awareness in HAT.'' We included studies that focused on the application of AI in HAT for tactical operations, explored the use of \ac{NLP} and reinforcement learning for improved communication, collaboration, and threat assessment, and addressed challenges related to trust, explainability, and ethical considerations. Furthermore, we included studies that explored the impact of AI-driven HAT on trust, explainability, and ethical considerations. We employed thematic analysis to identify key themes emerging from the reviewed literature, focusing on the opportunities and challenges associated with AI-driven HAT, with a particular emphasis on enhancing situational awareness, decision-making, and human-machine collaboration.

\vspace{1.0ex}

The remainder of this paper is organized as follows. Section \ref{motivation} discusses the integration of AI solutions into HAT. In Section \ref{delegated_autonomy_in_HAT}, we discuss the concept of delegated autonomy in HAT, exploring different levels and the balance between human decision-making and automated systems in teaming scenarios. Section \ref{characteristics_key_components} presents the key components and characteristics defining HAT systems. Next, Section \ref{applications_of_human-autonomy} identifies and discusses the practical applications of HAT, presenting real-world examples where HAT has proven advantageous. Section~\ref{economics_of_HAT} explores the economic aspects of AI integration in HAT. \ref{Situation_models_and_shared_situational_awareness} provides a detailed discussion of situation models and shared situational awareness in HAT. Section~\ref{roles_of_AI} outlines the specific roles and contributions of AI in enabling tactical autonomy in HAT, emphasizing its ability to enhance human decision-making. The opportunities and challenges associated with using \ac{AI} to enhance \ac{HAT} in tactical autonomy are discussed in Section~\ref{opportunities_and_challenges}. The design of user interfaces and interaction mechanisms for HAT systems in tactical autonomy settings is explored in Section \ref{interaction_and_interface_design}. Section \ref{proposed_framework} introduces a proposed framework for AI-driven HAT in tactical operations, describes the key components, and provides guidance for future research and development. Finally, Section \ref{practical_recommendations} provides practical recommendations for implementing and optimizing HAT systems. The paper concludes in Section \ref{conclusion} with indications for future work.

\section{Motivation}
\label{motivation}

In this section, we describe the motivation for the integration of \ac{AI} solutions within \ac{HAT}, highlighting their transformative impact on collaboration, communication, and coordination in dynamic and complex tactical environments.

HAT is a rapidly evolving field that seeks to combine the strengths of humans and autonomous systems to achieve common goals. In recent years, the convergence of \ac{AI} and \ac{HAT} has emerged as a paradigm-shifting approach with the potential to revolutionize decision-making, situational awareness, and operational efficiency in dynamic and complex tactical environments. In tactical autonomy, HAT revolutionizes how humans and machines work together in dynamic and complex environments. The integration of AI solutions into HAT offers a compelling avenue to enhance the strengths of both human operators and intelligent autonomous systems, which is promising to advance tactical autonomy. This paper underscores the significance of this integration and presents opportunities, challenges, and directions for future work. We envision a landscape in which the symbiotic relationship between humans and autonomous systems can reshape tactical decision-making, enhance situational awareness, and maximize operational efficiency. By focusing on its transformative impact, this paper sets the stage for a future where collaboration, communication, and coordination in dynamic and complex tactical environments can be elevated to new heights, ultimately contributing to safer and more successful mission outcomes.

\vspace{1.0ex}
\noindent \textbf{Collaboration}. At the core of AI-driven HAT lies a new era of collaboration that redefines the possibilities of human operators and intelligent autonomous systems working together~\cite{veitch2022systematic}. AI technologies serve as bridges that enhance collaboration by boosting human capabilities through data-driven insights and analytical power~\cite{ferguson2003autonomous}. By seamlessly integrating AI solutions into the decision-making process, HAT systems can leverage real-time data analysis, predictive analytics, and pattern recognition to provide human operators with a comprehensive and dynamic understanding of the tactical situation~\cite{schaefer2021human, ferguson2003autonomous, van2018human, schubert2018artificial}. This improved collaboration enables operators to make informed decisions more quickly, which is often critical in tactical environments where split-second choices can impact mission success~\cite{schaefer2021human, van2018human, schubert2018artificial}.

\vspace{1.0ex}
\noindent \textbf{Communication}. Effective teamwork relies on empowering communication, and within the context of tactical autonomy, the integration of AI introduces a new dimension to this foundational aspect~\cite{funke2022teamwork}. \ac{NLP} and intelligent communication interfaces enable HAT systems to facilitate seamless interactions between humans and autonomous agents. Conversational AI, chatbots~\cite{mctear2022conversational}, and language translation tools enable real-time communication~\cite{meszaros2018trusted}, transcending language barriers and fostering a more inclusive and collaborative environment. This enhanced communication enables operators to convey complex instructions, receive real-time updates, and seek clarifications, thus simplifying the decision-making processes and reducing ambiguity in high-stress scenarios. As described by Shively et al.~\cite{shively2018human}, HAT also incorporates a bi-directional communication approach, which transforms automation from a tool to a teammate. This dynamic communication enables collaborative problem-solving, enabling seamless interactions and joint decision-making between automated systems and human operators.

\vspace{1.0ex}
\noindent \textbf{Coordination}. In dynamic and complex tactical environments, humans and autonomous systems must effectively coordinate their actions effectively~\cite{demir2018team}. Precise coordination in dynamic and complex tactical environments requires a level of precision that traditional approaches often struggle to achieve. AI-driven HAT transforms coordination into a finely tuned orchestration of human and autonomous actions. Autonomous agents equipped with reinforcement learning and multi-agent systems can execute tasks with adaptability and accuracy, aligning their actions with human operator intentions. This coordination optimizes resource allocation, minimizes response times, and ensures that tasks are executed efficiently, even in the face of unforeseen challenges. The result is a synchronized team that capitalizes on each member's strengths and operates in harmony to achieve the mission objectives.

\section{Delegated Autonomy in HAT}
\label{delegated_autonomy_in_HAT}

As technology advances, the integration of autonomy into various domains has become more prevalent~\cite{peeters2021hybrid}. Delegated autonomy, which is a critical concept in HAT, entails granting autonomous systems a certain level of decision-making authority while maintaining human oversight based on predefined rules, constraints, or algorithms~\cite{van2021delegation}. The degree of autonomy granted to machines can vary based on task complexity, system capabilities, and the context of the operation. Humans retain the ability to intervene, monitor, and override autonomous decisions when necessary, thus ensuring accountability and preventing potential errors. Delegated autonomy generally refers to the ability or a situation in which a human operator dynamically assigns certain tasks or responsibilities to an autonomous system, thereby allowing the system to operate independently within specified constraints~\cite{castelfranchi1998towards}. This can be achieved in various ways, depending on the specific tasks or responsibilities being delegated. The following are some practical examples of delegated autonomy.

\vspace{1.0ex}
\noindent \textbf{\ac{UAVs}}. In the field of aviation, \ac{UAVs} often operate with delegated autonomy. Autonomous drones can follow pre-planned flight paths, avoid obstacles, and adapt to changing weather conditions, while human operators maintain the authority to intervene in situations that require human judgment~\cite{cummings2011impact}. Studies have shown that \ac{HAT} systems can perform effectively in unmanned settings for search and rescue~\cite{alotaibi2019lsar, scherer2015autonomous}, infrastructure inspection~\cite{shakhatreh2019unmanned, lattanzi2017review}, and agriculture and traffic monitoring~\cite{menouar2017uav, huang2013development}.

\vspace{1.0ex}
\noindent \textbf{Robotic Systems}. Robots are being used in various industries, from manufacturing~\cite{pearce2018optimizing, li2023proactive} to healthcare~\cite{okamura2010medical}. In the context of robotic systems, surgical robots exemplify delegated autonomy in healthcare and other domains~\cite{lei2022human}. In the context of medical robots, a human surgeon delegates some of their autonomy to the robot, allowing it to perform certain tasks without direct human intervention~\cite{hussain2014use, howe1999robotics}. Surgeons control robotic arms to perform precise movements during surgeries, while the system's autonomy assists in error correction and stabilizing movements. Some of the benefits of medical robots include increased precision and accuracy~\cite{davies2000review, howe1999robotics}, enhanced efficiency~\cite{howe1999robotics}, minimized human error~\cite{howe1999robotics}, remote surgery and telemedicine~\cite{howe1999robotics}, \etc

\vspace{1.0ex}
\noindent \textbf{Autonomous Vehicles}. Self-driving cars operate with varying degrees of delegated autonomy~\cite{wang2020decision}. A vehicle's autonomous systems handle tasks like lane-keeping~\cite{blaschke2009driver, saito2016driver} and adaptive steering control~\cite{tada2016simultaneous, wang2016human, wang2015human, ercan2018predictive}, while the human driver remains responsible for monitoring the environment and taking control when needed~\cite{wang2020decision}.

\section{Key components and characteristics of HAT}
\label{characteristics_key_components}

Understanding the key components and characteristics of Human-Autonomy Teaming (\ac{HAT}) is important for exploring its wide-ranging applications, as discussed in Section~\ref{applications_of_human-autonomy}.

\subsection{Key Components of HAT}

Based on~\cite{roth2019function, sarter2023contributions, tokadli2022autonomy, li2023proactive, funke2022teamwork}, we identify the essential components and relevant aspects of integrating \ac{HAT} in practical contexts. These components and aspects guide the understanding and implementation of human-autonomy interaction and teaming, thus providing suitable methodologies for conducting experiments~\cite{el2023joint}.

\vspace{1.0ex}
\noindent \textbf{Human Operators}. The human component of \ac{HAT} consists of competent skilled individuals with the necessary expertise, decision-making abilities, and interpersonal communication skills to achieve team goals~\cite{sarter2023contributions}. Human workers engage in tasks requiring judgment, decision-making, creativity, and interpersonal communication~\cite{el2023joint}.

\vspace{1.0ex}
\noindent \textbf{AI}. In \ac{HAT}, \ac{AI} plays a crucial role in augmenting human abilities and driving team performance by providing cognitive capabilities, such as perception, reasoning, and decision-making, which enable autonomous systems to operate effectively in complex environments~\cite{roth2019function}. Careful design and integration of AI algorithms are essential for the reliability, credibility, and transparency of HAT systems. For more details, refer to Section~\ref{roles_of_AI}.

\vspace{1.0ex}
\noindent \textbf{Autonomous Systems}. This aspect of \ac{HAT} involves machinery, computer systems, or AI that can automate tasks and make predictions through AI algorithms~\cite{funke2022teamwork}. Autonomous systems enhance human abilities, enabling them to focus on complex tasks and decision-making.

\vspace{1.0ex}
\noindent \textbf{Interfacing with Autonomous Systems}. Communication plays a vital role in HAT. An ontology-based communication language allows direct interactions between AI and autonomous systems. Effective communication in HAT is facilitated by a communication language ontology and domain ontologies~\cite{el2023joint}. These ontologies ensure seamless communication between humans, AI, and autonomous systems, thereby enhancing collaboration and data exchange~\cite{el2023joint}.

\subsection{Characteristics of HAT}

Research on \ac{HAT} underscores the importance of team performance outcomes, collaboration processes, and effective training methods~\cite{schelble2023investigating,li2023proactive}. Some of the main characteristics of \ac{HAT} include:

\vspace{1.0ex}
\noindent \textbf{Heterogeneity}. HAT teams comprise diverse members with specific roles, and they leverage the strengths of AI systems to realize tasks that align with their capabilities~\cite{roth2019function}.

\vspace{1.0ex}
\noindent \textbf{Shared Cognition}. Developing shared mental models promotes effective teamwork within HAT and enhances team understanding and performance~\cite{schelble2023investigating}. This practice facilitates a deeper understanding of teammates' capabilities, limitations, objectives, and performance, thereby significantly facilitating efficient team processes and overall team performance. Moreover, developing shared mental models contributes to the establishment of shared situational awareness within teams~\cite{schelble2023investigating}.

\vspace{1.0ex}
\noindent \textbf{Collaboration and Communication}. Successful teamwork in HAT requires efficient collaboration and communication among humans, AI algorithms, and autonomous systems~\cite{funke2022teamwork}.

\vspace{1.0ex}
\noindent \textbf{Social Intelligence}. Leveraging social intelligence enhances the effectiveness of human team members' effectiveness in \ac{HAT}, enabling team members to effectively understand and support teammates effectively~\cite{li2023proactive}.

\section{Applications of HAT}
\label{applications_of_human-autonomy}

\begin{figure}[!h]
 \includegraphics[width=\linewidth]{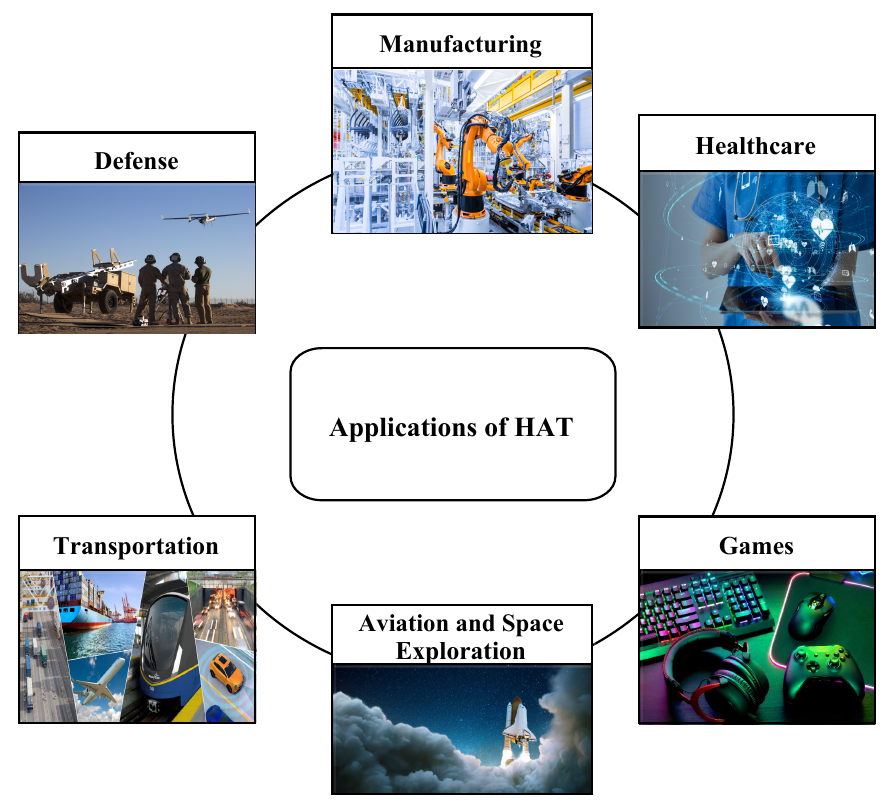}
 \caption{Applications of HAT.}
 \label{Fig:HAT}
\end{figure}

After identifying the essential building blocks of HAT in the previous section, we explore how HAT applications are revolutionizing various industries. HAT technology has the potential to revolutionize many industries. \ac{HAT} systems leverage the strengths of humans and autonomous systems to perform tasks with greater accuracy, speed, and reliability~\cite{lee2004trust}. These systems are increasingly being employed across various industries to exploit the strengths of humans and intelligent autonomous systems. These teams can improve safety, efficiency, and productivity across various domains. Figure~\ref{Fig:HAT} shows applications of \ac{HAT} in modern life.

\vspace{1.0ex}
\noindent \textbf{Defense}. In modern military applications, \ac{HAT} systems enable the seamless integration of human intelligence and strategic thinking with the speed, precision, and endurance of intelligent autonomous systems~\cite{chen2018human, schaefer2021human}. This integration enhances situational awareness, mission effectiveness, and operational efficiency~\cite{chen2018human, schaefer2021human}. By combining the strengths of humans and intelligent machines, \ac{AI} has the potential to revolutionize military operations and make the world a safer place.

\vspace{1.0ex}
\noindent \textbf{Manufacturing}. \ac{HAT} can optimize industrial processes by combining human expertise with automation~\cite{schelble2020towards}. Humans have cognitive abilities, problem-solving skills, and adaptability, and intelligent autonomous systems offer precision, strength, and speed~\cite{li2023proactive}. 

\vspace{1.0ex}
\noindent \textbf{Healthcare}. \ac{HAT} systems have the potential to revolutionize healthcare by assisting medical professionals in their work~\cite{hughes2016saving}. Modern \ac{HAT} systems can be used to analyze medical images, such as X-rays and \ac{MRI} scans, to identify signs of disease. HAT systems can also be used to analyze patient data, such as blood test results and medical history, to help physicians make more accurate diagnoses.

\vspace{1.0ex}
\noindent \textbf{Games}. \ac{HAT} principles can be applied to gaming to enhance player experiences by assisting with tasks that are difficult or time-consuming for humans, creating more engaging gameplay, and exploring new ways of interaction between players and autonomous systems in a virtual environment~\cite{izumigawa2020building, lucero2020human}.

\vspace{1.0ex}
\noindent \textbf{Aviation and Space Exploration}. \ac{HAT} is a promising technology with the potential to revolutionize aviation and space exploration~\cite{sarter2023contributions}. In aviation, cockpit automation involves collaboration between pilots and autonomous systems to safely operate aircraft~\cite{strybel2018effectiveness, r2016application}.

\vspace{1.0ex}
\noindent \textbf{Transportation}. \ac{HAT} can be applied to autonomous vehicles for passenger transportation and logistics. This involves collaboration between self-driving vehicles, human drivers, autonomous vehicles, and pedestrians in urban environments. In addition, HAT is considered essential for safe and efficient operation in the context of \ac{UAM} systems. \ac{UAM} is an emerging concept that refers to using aerial vehicles, such as drones or small electric aircraft, to transport people and goods within urban environments~\cite{goyal2018urban}. \ac{HAT} can play a crucial role in ensuring safe, efficient, and integrated operations within \ac{UAM} systems.

\section{The Economics of Integrating AI and HAT}
\label{economics_of_HAT}

Integrating \ac{AI} and \ac{HAT} for tactical autonomy brings about a range of economic benefits, particularly in domains where rapid and effective decision-making, enhanced situational awareness, and optimized resource utilization are critical. Although the potential of HAT applications across various industries is undeniable (Section~\ref{applications_of_human-autonomy}), a closer look at the economic impact of integrating AI and HAT is also important. This Section provides a detailed analysis of the potential economic benefits and challenges associated with such transformative collaboration. This study also examines the impact of AI-driven HAT on productivity, labor markets, cost-effectiveness, and overall economic growth. Additionally, the potential economic challenges and opportunities presented by the widespread adoption of HAT in various industries are highlighted.

\subsection{Economic Benefits of Integrating AI and HAT}

Several potential economic benefits of \ac{HAT} are discussed in prior studies~\cite{pazouki2018investigation, parasuraman2017humans}. These benefits include enhanced productivity, safety, operational efficiency, and cost reduction. \ac{HAT} enables human workers to concentrate on creative and strategic tasks.

\vspace{1.0ex}
\noindent \textbf{Improved productivity}. Effective collaboration between humans and autonomous systems can significantly improve productivity. Autonomous systems excel at performing repetitive tasks at high speed and accuracy, whereas humans contribute their expertise and decision-making skills, which are lacking in autonomous systems.

\vspace{1.0ex}
\noindent \textbf{Reduced Operational Costs}. Incorporating AI and HAT in tactical autonomy not only enhances the efficiency and effectiveness of critical missions but also contributes to substantial economic gains through reduced costs, increased success rates, and optimized resource utilization. Automation leads to cost reduction by delegating routine tasks, such as manufacturing, transportation, and customer service, to autonomous systems. This, in turn, allows human workers to concentrate on more creative and strategic tasks that demand problem-solving skills and creativity. Hence, automating routine and data-intensive tasks through AI-powered autonomous systems reduces labor costs and minimizes human intervention. These optimizations translate to significant operational cost savings.

\vspace{1.0ex}
\noindent \textbf{Improved Situational Awareness}. As explained below, AI enhances the information available to human operators, providing a comprehensive view of the tactical environment. This leads to better-informed decisions and minimizes the financial consequences of inadequate awareness.

\vspace{1.0ex}
\noindent \textbf{Scalability and Flexibility}. AI-driven intelligent autonomous systems can adapt to changing tactical conditions and complex requirements, enabling scalable operations without increasing human labor costs~\cite{schelble2020towards}.

\subsection{Economic Challenges of Integrating AI and HAT}

In addition to economic benefits, \ac{HAT} presents potential economic challenges. One such challenge, for example, is job displacement because autonomous systems take over tasks currently performed by humans. In addition, increased reliance on autonomous systems can pose safety risks if they are not properly designed and operated. Here, we discuss some potential challenges in detail.

\vspace{1.0ex}
\noindent \textbf{Job displacement}. Automation could lead to job displacement because modern and intelligent autonomous systems can take over tasks currently performed by humans. This shift might hurt the economy, resulting in higher unemployment rates and lower wages.

\vspace{1.0ex}
\noindent \textbf{Increased safety risks}. \ac{HAT} could lead to increased safety risks if autonomous systems are not properly designed and operated. For example, mistakes made by autonomous systems can result in accidents or injury.

\vspace{1.0ex}
\noindent \textbf{Privacy concerns}. Automation could raise privacy concerns because autonomous systems can collect and store large amounts of data about human users. These data could be used for marketing or other purposes without the user's consent.

\section{Situation Models and Shared Situational Awareness in HAT}
\label{Situation_models_and_shared_situational_awareness}

The economic analysis in Section~\ref{economics_of_HAT} highlighted the importance of efficient decision-making in HAT. However, effective decision-making requires a shared understanding of the situation. This section explores how situation models and shared situational awareness facilitate the flow of information required for human and autonomous systems to work together seamlessly.

\subsection{Situation Models of HAT}

\begin{figure*}[!h]
 \includegraphics[width=\linewidth]{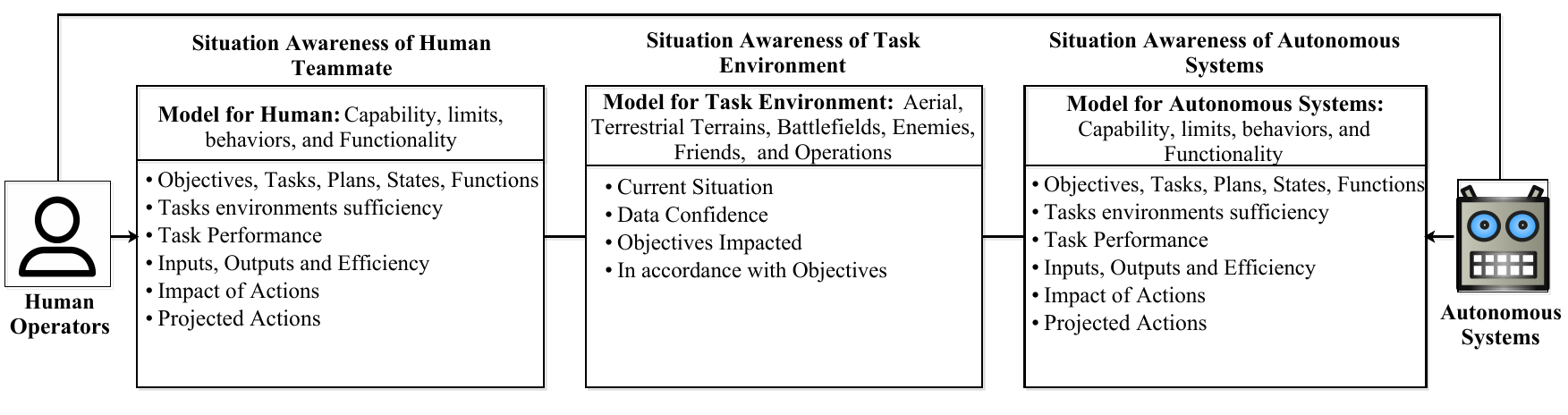}
 \caption{Situation models for HAT~\cite{el2023joint}.}
 \label{Fig:SA}
\end{figure*}

In the context of \ac{HAT}, situation models and shared situational awareness play a crucial role in ensuring effective collaboration between humans and autonomous systems in complex environments~\cite{onken1997cockpit}. Situation models represent an individual's internal understanding of the world, their experiences, and others. This understanding is dynamic and constantly updated based on sensory inputs and mental models (see Figure~\ref{Fig:SA}).  For effective collaboration in complex environments, HAT requires humans, AI, and autonomous systems to develop internal situation models. Here, are the key aspects of the situation models in HAT.

\vspace{1.0ex}
\noindent \textbf{Situation}. Similar to how humans in \ac{MDO} rely on situational awareness, AI algorithms must develop and maintain an accurate model of the world. This is crucial for informed decision-making\cite{onken1997cockpit, national2021human, jones2011using, kokar2012situation}. Effective \ac{HAT} systems emphasize shared mental models and team situational awareness. Advanced AI techniques can improve these aspects by providing humans with insights into the decision-making processes of machine learning models, further enhancing situational awareness and shared understanding~\cite{paleja2021utility}.

\vspace{1.0ex}
\noindent \textbf{Task Environment}. A dynamic function allocation mechanism has been proposed for future \ac{HAT} systems, where tasks are distributed among human and autonomous teammates based on their capabilities \cite{roth2019function}. This requires an updated model of the work environment, including current goals, assignments, plans, and the state of humans and automation involved~\cite{endsley2017here}.

\vspace{1.0ex}
\noindent \textbf{Teammate awareness}. As humans must understand AI reliability, AI may require a model of the current state of its human teammates to perform assigned tasks effectively \cite{xu2023transitioning, carroll2019utility}.

\vspace{1.0ex}
\noindent \textbf{Self-awareness}. Being aware of one's capabilities is important. Team members who recognize fatigue, workload, or inadequate training can shift tasks to optimize performance \cite{dierdorff2019power, dorneich2017evaluation}. AI may need to develop a model of its performance limitations to indicate when human intervention is needed or whether its calculations are accurate \cite{national2021human, chella2020developing}.

These individual situation models are important for achieving shared situational awareness, which is the focus of the following subsection.

\subsection{Shared Situational Awareness}

Situational awareness in \ac{HAT} refers to individuals' and teams' ability to perceive, understand, and anticipate relevant information and events in their operational environment~\cite{nofi2000defining, harrald2007shared}. This extends beyond human team members to include collaborative autonomous systems and robots. Developing shared situational awareness requires collaboration between humans, autonomous systems, and AI algorithms. Humans rely on sensory inputs such as vision and hearing, and autonomous systems rely on sensor data and AI algorithms for safety~\cite{nofi2000defining, harrald2007shared}. Effective communication, information sharing, and collaboration are crucial for maintaining shared awareness over time in HAT. The following are some challenges to achieving shared situational awareness:

\vspace{1.0ex}
\noindent \textbf{Information Overload}. The vast amount of data generated by autonomous systems can overwhelm human agents~\cite{nofi2000defining, salmon2017distributed}. This makes it harder for users to concentrate on important tasks. HAT systems should provide mechanisms to filter and prioritize information, ensuring that human agents only receive relevant and actionable data~\cite{salmon2017distributed}.

\vspace{1.0ex}
\noindent \textbf{Cognitive Overload}. Processing large amounts of information while making decisions can lead to cognitive overload in human agents \cite{nofi2000defining}.  HAT systems can mitigate this by incorporating intelligent algorithms to facilitate data analysis and decision-making, which reduces the cognitive burden of humans.

\vspace{1.0ex}
\noindent \textbf{Training and Interfaces}. Effective use of tools and interfaces to facilitate shared situational awareness is crucial for human agents in HAT scenarios~\cite{salas2017situation, stout2017role}.  These tools include dashboards, augmented reality displays, and communication systems that present relevant information.  Training programs should cover not only technical aspects but also emphasize human-autonomy collaboration strategies and best practices. User-friendly interfaces and well-designed human-autonomy interaction mechanisms can also help reduce the learning curve and improve usability.

In summary, building a common operating picture through situation models and shared situational awareness is essential for effective HAT. By carefully managing information flow, providing appropriate training, and using well-designed interfaces, we can ensure successful collaboration between humans and autonomous systems.

\section{Roles of AI in HAT for Tactical Operations}
\label{roles_of_AI}

Section~\ref{Situation_models_and_shared_situational_awareness} highlighted the importance of shared understanding in HAT. However, effectively processing the enormous amount of information required by such models is a critical challenge. This section explores how AI capabilities are leveraged in HAT to address this challenge, with a specific focus on tactical operations.

\ac{AI} has created a revolutionary transformation in numerous domains and industries, introducing advanced capabilities previously beyond reach. In the medical and healthcare fields, for example, \ac{AI} plays a pivotal role in disease diagnosis, treatment development, and personalized patient care~\cite{yeasmin2019benefits}. Diagnostic AI-powered systems exhibit remarkable accuracy in interpreting medical images, such as X-ray and MRI images, thereby helping medical professionals make more informed decisions. These systems excel at analyzing complex medical images and identifying tumors or anomalies that human professionals may overlook. Additionally, AI contributes to drug and therapy innovation, allowing personalized treatment plans for individual patients~\cite{yeasmin2019benefits}. In the digital marketing context, AI is invaluable because it enhances the precision of targeted advertising, elevates customer experiences, and predicts consumer behavior~\cite{mitic2019benefits, jain2020transforming}.

In the context of tactical autonomy, the potential roles of \ac{AI} in \ac{HAT} become even more critical. Tactical autonomy refers to the ability of autonomous systems to make decisions and take action in real-time, often in complex and unpredictable environments~\cite{hagos2022recent}. AI-powered systems can provide soldiers with real-time battlefield information and can also be used to develop autonomous weapons systems that can operate without human intervention~\cite{hagos2022recent}. The potential benefits of \ac{AI} in tactical autonomy are significant. AI systems can help improve situational awareness, identify and track targets, plan and execute missions, communicate with other systems, and make decisions. This can lead to increased safety, efficiency, and effectiveness in military operations~\cite{schaefer2021human, vorm2020computer}.

\vspace{1.0ex}
\noindent \textbf{Provide situational awareness}. As explained above, \ac{AI} systems can collect, process, and analyze large amounts of data from sensors, cameras, and other sources to provide humans with an enhanced understanding of complex and dynamic environments~\cite{simpson2021real}.  It offers real-time insights, predictions, and suggestions that help human operators make informed decisions. This can help humans make better decisions and take appropriate actions.

\vspace{1.0ex}
\noindent \textbf{Plan and execute missions}. AI systems can plan and execute missions considering various factors, such as the team environment, capabilities, and risks. This allows human operators to make high-level decisions and manage the overall mission.

\vspace{1.0ex}
\noindent \textbf{Communicate with other systems}. AI systems can communicate with other systems, such as other autonomous vehicles and command and control centers. This will help ensure that the team works effectively together.

\vspace{1.0ex}
\noindent \textbf{Learn and adapt}. AI systems can learn from experience and adapt their behavior accordingly. This allows them to become more effective over time.

\vspace{1.0ex}
\noindent \textbf{Make decisions}. AI systems can make decisions that are enhanced even in complex and uncertain situations. This can help humans avoid making mistakes and ensure that the team always acts in the best interests.

However, some challenges must be addressed. One challenge is to ensure that AI systems are reliable and safe. Another challenge is the need to develop trust between humans and AI systems. Finally, there are ethical considerations that must be considered, such as the potential for AI systems to be used for malicious purposes. Despite these challenges, \ac{AI} can revolutionize tactical autonomy. As \ac{AI} systems continue to develop, we expect to see even more innovative and effective applications of \ac{AI} in this area.

\section{Human-AI Interaction in HAT for Tactical Autonomy}
\label{interaction_and_interface_design}

Effective human-AI interaction is crucial for successful HAT implementation. HAT systems are designed to allow human and autonomous systems to work together effectively in complex and dynamic environments. Section~\ref{roles_of_AI} explored how AI empowers HAT systems, particularly tactical operations. However, to harness the full potential of AI and collaborate effectively, well-designed user interfaces are essential.  This section explores the key factors and \ac{HMI} design principles to consider when designing \ac{HAT} interfaces, ultimately ensuring seamless human-machine collaboration. For example:

\vspace{1.0ex}

\noindent \textbf{Transparency and Explainability}. These principles are fundamental to ensure that human operators can understand, trust, and effectively collaborate with autonomous systems~\cite{chen2016human, endsley2023supporting}. Transparency refers to the extent to which the human operator can understand how an autonomous system works and why it makes the decisions it does~\cite{chen2016human, robbemond2022understanding}. In the context of HAT interfaces, transparency involves providing operators with insights into the functioning of an autonomous system and making decisions~\cite{paleja2021utility}. In contrast, explainability refers to the extent to which a human operator can understand the rationale behind an autonomous system's decisions~\cite{samek2019explainable, ehsan2021expanding}. Achieving real-time transparency and explainability in HAT interfaces is important for human operators to rely on AI-driven recommendations and actions. However, achieving explainability and transparency in human-AI interaction within the context of tactical autonomy settings requires several advanced strategies and technologies~\cite{guidotti2018survey}.

\vspace{1.0ex}

\noindent \textbf{Context Awareness}. HAT interfaces should provide human operators with the status of autonomous agents, the overall mission, and a clear and real-time understanding of the underlying tactical environment~\cite{alix2021empowering, dubey2020haco}. Based on the current task and context, displaying relevant real-time information about the tactical environment, such as maps, sensor data, and mission objectives, is important to enhance human operators' situational awareness~\cite{endsley1995toward, lafond2014hci}.

\vspace{1.0ex}

\noindent \textbf{Adaptability and Flexibility}. Tactical autonomy settings are often characterized by rapidly changing conditions. HAT systems are designed to operate in dynamic and unpredictable environments where the conditions can change rapidly. Therefore, the human-AI interaction mechanism must be able to adapt to these changes and be sufficiently flexible to accommodate various tasks, goals, and environments~\cite{miller2007designing}. An adaptable HAI system can adjust its behavior, decision-making processes, and responses to accommodate these changes~\cite{erol2016tangible}. This allows the system to remain effective and relevant in a constantly evolving context.

\vspace{1.0ex}

\noindent \textbf{Safety and Redundancy}. These aspects help improve the reliability, robustness, and trustworthiness of HAT systems in dynamic and potentially dangerous environments~\cite{li2023trustworthy, zaki2021reliability}. Safety refers to measures and mechanisms implemented to prevent accidents, mitigate risks, and ensure that the system operates without harming humans, property, or the mission. Redundancy, on the other hand, involves duplicating critical components or functions within the HAT system to ensure that backups are available in case of failure~\cite{ramos2019autonomous}. The work in~\cite{alexander2018state} provides an overview of the current state of safety solutions and challenges in ensuring the safety of autonomous systems.

\vspace{1.0ex}

\noindent \textbf{Shared Mental Model}. Developing shared mental models between humans and autonomous systems is important for effective collaboration and communication~\cite{andrews2023role, bansal2019beyond}. A shared mental model refers to a common understanding of the task, environment, and capabilities of humans and autonomous systems~\cite{kaur2019building, lee2004trust}. To promote the development of shared mental models and strengthen the collaboration between humans and autonomous systems, several critical strategies should be implemented~\cite{tokadli2022autonomy}. First, clear communication is very important. This involves using a common language or terminology that is understandable to humans and autonomous systems. Second, feedback mechanisms play a significant role. Providing humans with feedback on the performance of autonomous systems enhances understanding and trust~\cite{tokadli2022autonomy}. Third, visualization tools are also equally important. They help humans understand the internal states and reasoning processes of autonomous systems. Finally, designing a human-autonomy system user interface and interaction mechanisms to facilitate effective communication and collaboration is important~\cite{tokadli2022autonomy, schelble2020designing}.

\section{Opportunities and Challenges}
\label{opportunities_and_challenges}

Moving from design principles to practical applications requires careful consideration of the broader landscape. This section explores the opportunities and challenges associated with using AI to enhance HAT in tactical autonomy.

\subsection{Opportunities of HAT in Tactical Autonomy}

\ac{HAT} is a promising technology with the potential to improve the safety and effectiveness of military mission-critical operations~\cite{huang2020human}. As discussed above, \ac{HAT} is a concept that focuses on the collaboration and interaction between humans and autonomous systems or AI-driven technologies, particularly in scenarios where both entities work together toward a common goal~\cite{shively2018human}. This concept presents numerous opportunities across various domains. It is particularly important in the context of tactical autonomy and in defense and military areas for several compelling opportunities:

\vspace{1.0ex}
\noindent \textbf{Enhanced Decision-making}. Autonomous systems powered by \ac{AI} can analyze vast amounts of data rapidly, providing human operators with real-time, data-driven insights to make more informed and effective decisions.

\vspace{1.0ex}
\noindent \textbf{Enhanced performance}. \ac{HAT} can significantly enhance overall performance by leveraging the strengths of humans and intelligent autonomous systems~\cite{cummings2014man, metzger2005automation, lyons2021human}. Humans possess cognitive abilities such as creativity, intuition, and complex decision-making, whereas autonomous systems provide computational power, precision, and efficiency.

\vspace{1.0ex}
\noindent \textbf{Increased efficiency}. \ac{HAT} systems can help reduce the risk of human error by automating tasks that are prone to human error. This can lead to safer operations, fewer accidents, and improved efficiency~\cite{richards2020measure, lyons2021human}. In addition to this, \ac{HAT} systems can help human operators focus on more important tasks. For example, in the aviation industry, \ac{HAT} systems can automate tasks such as aircraft system monitoring and navigation. This allows the human pilot to focus on critical tasks such as decision-making and communication with air traffic control~\cite{billings1996human}.

\vspace{1.0ex}
\noindent \textbf{Improved safety}. \ac{HAT} systems can help improve safety by automating complex tasks that are extremely difficult for humans to perform~\cite{lyons2021human}. For example, autonomous systems can be used to perform difficult tasks or pose risks to humans, such as driving vehicles under hazardous conditions. Autonomous systems can also be programmed to follow safety procedures more consistently than humans, and they can be equipped with sensors that detect hazards that humans may not be able to detect.

\vspace{1.0ex}
\noindent \textbf{Risk reduction}. By integrating intelligent autonomous system agents, human team members can delegate high-risk tasks to autonomous agents, which mitigates risks and improves overall safety in dynamic and complex environments~\cite{richards2020measure}.

\vspace{1.0ex}
\noindent \textbf{Enhanced Capabilities}. \ac{HAT} can help enhance the capabilities of human operators by providing them with access to information and resources that they would not otherwise have~\cite{arkin2009governing}. This can help them make better decisions and take more effective actions.

\vspace{1.0ex}
\noindent \textbf{Adaptability and Flexibility}. HAT systems should be flexible and adaptable to dynamic environments. Both humans and autonomous systems should adjust their behaviors and decision-making in response to evolving situations. Autonomous systems can adapt to dynamic and changing environments more rapidly than humans~\cite{boardman2019exploration, lyons2021human}. Their ability to process real-time data and adjust their actions accordingly enhances the team's overall adaptability and resilience.

\subsection{HAT Challenges in Tactical Operations}

While \ac{HAT} offers significant opportunities, it also presents critical challenges that must be addressed for successful implementation. Here, we present some key challenges associated with \ac{HAT} for tactical autonomy:

\vspace{1.0ex}
\noindent \textbf{Trust}. Investigating the factors that influence human trust in AI systems and developing strategies to enhance trust between humans and \ac{AI} is crucial. Trust is critical for effective collaboration and decision-making in human-autonomy teams because humans must be confident that AI systems will behave safely and predictably. Therefore, AI systems should be designed to be transparent and explainable so that humans can understand how they make decisions. For example, the authors in~\cite{mcneese2019understanding} emphasize the importance of exploring trust dynamics within human-autonomy teams. They suggest a need for a detailed and qualitative analysis of team processes to understand how trust can be established or eroded over time. Such insights can contribute to more refined human-autonomy team designs and guide the development of autonomous agents that prioritize the element of trust.

\vspace{1.0ex}
\noindent \textbf{Reliability}. It is important to ensure that AI-powered autonomous systems are reliable~\cite{lee2004trust}. Although HAT offers several potential benefits, ensuring the reliability and trustworthiness of these systems is essential for maintaining safety, ethical standards, and public confidence~\cite{lee2004trust}. In the context of this study, trustworthiness refers to how well an autonomous agent earns the trust of other agents in the team, including humans and other autonomous agents. Trustworthiness is important because it allows humans to trust autonomous agents to operate safely and reliably. One of the key challenges in \ac{HAT} is ensuring that humans and autonomous systems can trust and cooperate. This is especially important in complex and dynamic environments, where human and autonomous systems must be able to make quick decisions and adapt to changing conditions. Researchers must develop methods for improving the transparency and explainability of autonomous systems and methods for training humans to better understand and work with autonomous systems.

\vspace{1.0ex}
\noindent \textbf{Lack of Transparency and Explainability of AI}. Ensuring real-time transparency and explainability is crucial for building trust and enhancing situational awareness in \ac{AI} systems~\cite{balasubramaniam2023transparency, larsson2020transparency}. Human understanding of autonomous system decision-making is essential for effective collaboration, particularly in situations with significant consequences such as self-driving cars or autonomous weapons systems. However, designing transparent algorithms and interfaces to interpret autonomous system actions is a complex challenge.

\vspace{1.0ex}
\noindent \textbf{Human-machine Collaboration}. It is important to ensure that humans and autonomous systems can effectively work together. This requires careful design of the \ac{HAT} system, and training for both humans and autonomous systems. Hence, the need for better human-machine interfaces that allow humans and machines to work together effectively is a critical challenge.

\vspace{1.0ex}
\noindent \textbf{Shared Situational Awareness}. Effective collaboration between humans and intelligent autonomous system agents requires maintaining shared situational awareness among the team members~\cite{onken1997cockpit}. However, the critical challenge lies in designing and ensuring that all members have access to relevant information and can interpret and understand it consistently. Therefore, shared situational awareness in HAT is critical for human agents. This is because human agents can be overwhelmed by the information provided by autonomous systems. Processing a large amount of information while making critical decisions can lead to cognitive overload in human agents, and human agents may also need to be trained to effectively use tools and interfaces that facilitate shared situational awareness.

\vspace{1.0ex}
\noindent \textbf{Workload Distribution}. Allocating workloads appropriately between humans and autonomous systems is crucial to prevent cognitive overload or underutilization of computing resources. Achieving a balance that optimizes the strengths of both team members and guarantees efficient task execution is a critical challenge, particularly in complex and dynamic environments.

\vspace{1.0ex}
\noindent \textbf{Ethical Implications}. \ac{HAT} raises several ethical implications, such as the potential for autonomous systems can make decisions that result in harm to humans. As \ac{HAT} systems become more sophisticated, it is important to consider ethical implications carefully prior to deploying HAT systems. For example, how can we ensure that \ac{HAT} systems are used safely and responsibly not to harm humans (either intentionally or unintentionally)? How can we prevent HAT systems from being used for malicious purposes? How can we ensure that HAT systems are used in a fair and just manner? Researchers should work with policymakers and ethicists to develop ethical guidelines for the development and use of \ac{HAT} systems.

\subsection{Decision Logic of Autonomous Agents}
Understanding the decision logic of autonomous agents is essential for ensuring safe and effective collaboration between humans and intelligent autonomous systems. 

\vspace{1.0ex}
\noindent \textbf{Black-box Models}. Many autonomous systems use black-box models, which are \ac{ML} models that are trained on large amounts of data, where the relationships between the inputs and outputs of the models are complex and nonlinear. In the HAT context, black-box models can be used to control autonomous systems in various ways~\cite{hou2021impacts}. For example, a black-box model can be used to control the navigation of autonomous vehicles or the weapons system of autonomous robots. However, using black-box models in HAT poses several challenges. One challenge is that it can be difficult for humans to understand how autonomous systems are making decisions~\cite{asatiani2020challenges}. This can lead to a lack of trust in autonomous systems and make it difficult for humans to collaborate effectively with them.

\vspace{1.0ex}
\noindent \textbf{High-dimensionality of Data}.  Autonomous systems often process large amounts of high-dimensional data. This can make it difficult for humans to visualize data and understand the factors that influence an autonomous system's decisions~\cite{el2016mining}. When data are high-dimensional, this means that different features (variables) that are being measured~\cite{trunk1979problem}. In the HAT context, high-dimensional data can make it difficult for humans to understand how autonomous systems make decisions. This challenge pertains to the complexity and volume of data that autonomous systems generate and use, and it can impact various aspects of HAT, including decision-making, situational awareness, and communication between team members.

\vspace{1.0ex}
\noindent \textbf{Uncertainty of Data}.  Autonomous systems often operate in uncertain environments. This can make it difficult for humans to determine the reliability of autonomous system decisions. Uncertainty of data is a significant challenge in HAT and can impact the effectiveness, safety, and trustworthiness of human-autonomy systems. In HAT, uncertainty can arise from various sources and manifest in different forms, including noisy or incomplete sensor data, environmental uncertainty, and human-automation interaction uncertainty~\cite{olson2013exploration}. The data may also be outdated or may not represent the current situation. In addition, the data may be biased or manipulated by an adversary. Several methods can be used to address the challenge of uncertainty in HAT data~\cite{marathe2014confidence}. For example, it is important to use data that are as high quality as possible. It is also important to employ advanced techniques to detect and mitigate data uncertainty. Furthermore, it is important to design HAT systems that are robust against uncertainty.

\vspace{1.0ex}
\noindent \textbf{Unforeseen Biases}. Another critical challenge is the potential for unforeseen biases. AI systems are trained on data, and if the data are biased, the AI system will be biased as well. This could lead AI systems to make unfair or discriminatory decisions~\cite{eitel2021beyond, benjamins2021choices}. It is important to carefully select the data on which AI systems are trained and to employ techniques to mitigate bias. Autonomous systems may be biased in ways that are not immediately obvious to humans. This can lead to autonomous systems making poor decisions.

Addressing these critical challenges in integrating and deploying HAT in AI-powered domains is essential to maximize benefits and ensure safe, ethical, and effective operation in complex and dynamic environments. Despite these challenges, we believe that the potential benefits of using AI in HAT are immense. By combining the strengths of humans and machines, AI can revolutionize military operations and make the world a safer place.

\subsection{Function Allocation Challenges in HAT}
Function allocation refers to the distribution of tasks and responsibilities between humans and autonomous systems within a team~\cite{roth2019function}. Allocation is crucial for optimizing team performance, ensuring efficiency, and maintaining safety in complex operational environments. It determines how information is processed, integrated, and presented to human operators to deliver relevant and timely information that enhances situational awareness. This process includes considerations such as data fusion, visualization techniques, and feedback mechanisms from autonomous systems to humans~\cite{stanton2006distributed}. Traditionally, function allocation has been based on what humans and machines are good at. However, this approach overlooks several important factors that must be addressed. A more comprehensive approach involves analyzing task demands, exploring task distribution strategies, examining the interdependence between humans and machines, and considering the associated trade-offs~\cite{roth2019function}. The work in \cite{roth2019function} highlights several key considerations in function allocation that are particularly relevant in the era of human-autonomy teaming. Addressing these considerations can help human-autonomy teams leverage the strengths of both humans and machines, leading to improved performance, decision-making, and overall mission success in tactical operations. Some of these considerations relevant to this paper are discussed below.

\vspace{1.0ex}
\noindent \textbf{Cognitive Workload Distribution}.  A crucial consideration is the distribution of cognitive workload between humans and intelligent autonomous systems. Repetitive tasks, rule-based tasks, or processes that involve the rapid processing of large amounts of data are often better suited for automation. Conversely, tasks requiring creativity, complex decision-making based on contextual understanding, or ethical considerations are generally more suitable for human operators.

\vspace{1.0ex}
\noindent \textbf{Situational Awareness}. Function allocation plays a critical role in shaping situational awareness in tactical operations, particularly when integrating AI into \ac{HAT}. Situational awareness relies heavily on effective function allocation, which impacts the cognitive load experienced by human operators. By assigning tasks appropriately based on their complexity and cognitive demands, operators can enhance their ability to maintain situational awareness~\cite{endsley1995toward}. For example, automating routine and repetitive tasks allows human operators to focus on higher-level cognitive tasks that require situational understanding and decision-making~\cite{endsley1995toward}. Task allocation significantly influences the level of collaboration and interaction between humans and autonomous systems. Collaborative function allocation models, where humans and AI work together, can improve situational awareness by leveraging the strengths of both entities. This collaboration may involve shared decision-making, coordinated task execution, and continuous feedback loops~\cite{parasuraman2009adaptive}.

\vspace{1.0ex}
\noindent \textbf{Flexibility and Adaptability}. The allocation of functions should be flexible and adaptable to changing circumstances, such as dynamic situational awareness in tactical environments~\cite{demir2017team}. Strategies for function allocation that prioritize adaptability and flexibility enable quick adjustments to task assignments and information flow. This adaptability ensures that situational awareness remains robust even in evolving scenarios or unexpected events. Human-autonomy teams operate in dynamic environments where tasks and priorities may shift. Therefore, systems must be designed with the ability to reassign tasks or transfer control seamlessly based on real-time conditions and input from both humans and machines.

\vspace{1.0ex}
\noindent \textbf{Transparency and Trust}. Clear communication and transparency in function allocation are essential for building trust within the team. Human operators should understand how tasks are distributed among humans and autonomous systems, including the criteria used for decision-making. Transparent allocation enhances trust, reduces uncertainty, and promotes effective collaboration.

\vspace{1.0ex}
\noindent \textbf{Ethical and Legal Considerations}. Function allocation must also consider ethical and legal implications. This includes ensuring that humans retain control over critical decisions, addressing potential biases in algorithmic decision-making, and adhering to regulatory frameworks governing autonomous systems in specific domains such as defense or healthcare.

\begin{figure*}
\centering
\begin{tikzpicture}[mindmap, grow cyclic, every node/.style=concept, concept color=black!20, 
    level 1/.append style={level distance=4.5cm,sibling angle=90},
    level 2/.append style={level distance=3cm,sibling angle=45}, align=center]

\node{AI-Driven HAT in Tactical Operations}
    child { node {Trust and Transparency} 
        child { node {Explainable AI Models} }
        child { node {Trust Calibration Mechanisms} }
    }
    child { node {Function Allocation}
        child { node {Dynamic Task Distribution and Adaptability} }
        child { node {Cognitive Load Optimization} }
    }
    child { node {Situational Awareness}
        child { node {AI-Enhanced Perception} }
        child { node {Shared Mental Models} }
    }
    child { node {Ethical Considerations}
        child { node {Ethical Decision Support} }
        child { node {Accountability Frameworks} }
    };

\end{tikzpicture}
\caption{Proposed Framework for AI-Driven HAT in Tactical Operations}
\label{fig:HAT-framework}
\end{figure*}
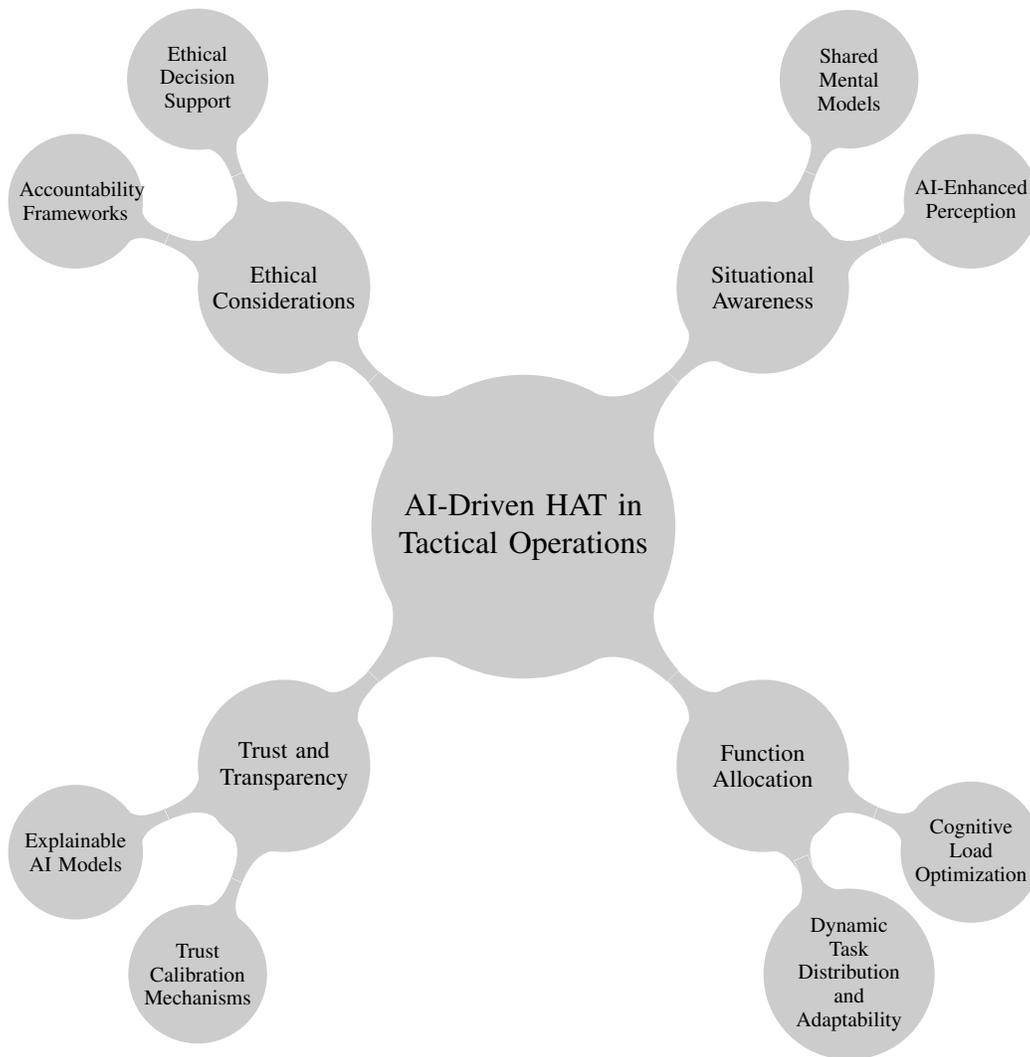

\section{Proposed framework for AI-Driven HAT in Tactical Operations}
\label{proposed_framework}

In this section, we present a comprehensive framework (Figure~\ref{fig:HAT-framework}) for AI-driven HAT in tactical operations, and we provide a conceptual structure to enhance the integration of AI into these environments. By identifying and organizing key elements, this framework can guide future research and development. It comprises four main components: trust and transparency, function allocation, situational awareness, and ethical considerations, each representing a critical aspect of AI-driven HAT that requires further exploration. By providing a structured approach, the proposed framework facilitates a systematic investigation of the challenges and opportunities associated with AI-driven HAT for tactical operations. We include comparative insights, new examples, and potential real-world implementation strategies to illustrate how our work advances beyond the existing literature. Future research is needed to develop, implement, and validate these concepts in real-world settings.

\subsection{Trust and Transparency}

\noindent \textbf{Explainable AI Models}. XAI models are crucial to ensure that human operators can understand AI decisions, particularly in time-sensitive environments. For example, in military drone operations, a tactical XAI model can provide real-time explanations for target identification using visual overlays to highlight specific features and auditory alerts to convey urgency. The proposed framework introduces a tactical XAI system that is tailored to the demands of such environments, thereby providing concise, actionable insights during mission-critical moments. It envisions a multi-modal explanation system that employs visual, auditory, and tactile cues to communicate AI reasoning without overwhelming the operator's cognitive load. In addition, an adaptive explanation component adjusts the depth and complexity of the information based on the operator's expertise and cognitive state, providing flexible levels of detail according to operational needs and time constraints.

\vspace{1.0ex}

\noindent \textbf{Trust Calibration Mechanisms}. In HAT systems, trust calibration must be adaptive rather than static to respond to evolving tactical scenarios. Drawing inspiration from reinforcement learning techniques, the proposed framework introduces a dynamic trust calibration system that adjusts AI autonomy based on operator behavior, AI performance, and mission criticality. In this system, trust levels are continuously recalibrated through real-time feedback to assess factors such as behavioral cues, physiological signals, and direct operator inputs. For example, in time-sensitive missions like search-and-rescue, if the AI system consistently performs with high accuracy, its autonomy is increased, which minimizes the need for human verification. Conversely, if the AI system misclassifies targets or poses risks, trust levels decrease, which increases human oversight. By autonomously adapting to operator inputs, mission outcomes, and evolving conditions, the underlying system enables seamless human-AI collaboration, enhancing decision-making through continuous real-time feedback and adjustments. This adaptive trust mechanism helps human operators remain confident in AI decisions, leading to better teamwork, especially during critical missions.

\subsection{Function Allocation}

\noindent \textbf{Dynamic Task Distribution and Adaptability}. Existing function allocation in HAT systems is primarily static and often fails to accommodate changing tactical conditions. To address this limitation, the proposed framework employs adaptive algorithms capable of dynamically distributing tasks based on variables, such as operator cognitive load, stress indicators, and mission requirements. The proposed framework emphasizes the importance of context-aware algorithms that modify task allocation in real-time, considering factors such as operator cognitive state, stress levels, and mission complexity. By considering variables such as mission type, environmental conditions, and operational data, these adaptive algorithms can optimize the distribution of responsibilities between humans and AI, ensuring that task allocation remains responsive to evolving tactical situations. Achieving this level of adaptability requires AI-driven HAT systems to be designed with flexibility as a core feature. AI architectures should be capable of rapid reconfiguration in response to new inputs, enabling them to adjust to shifting mission objectives, environmental changes, and variations in operator states. For example, reinforcement learning models can be employed to adapt task strategies based on real-time feedback, continuously optimizing system performance under changing conditions. Furthermore, incorporating modular AI components, such as plug-and-play sensors and dynamic data processing units, facilitates rapid updates, which ensures the system's resilience against unforeseen changes. This approach not only enhances AI efficiency but also guarantees better alignment with operators' immediate needs, thereby improving overall mission success rates and operator trust. For example, complex tasks like real-time threat assessment, can be assigned to AI systems during high-stress scenarios, whereas decision-making regarding ethical dilemmas, such as potential civilian casualties, remains the responsibility of human operators.

\vspace{1.0ex}

\noindent \textbf{Cognitive Load Optimization}. In traditional HAT systems, the cognitive load of human operators is often either underestimated or ignored, leading to reduced performance and potential mission failures. Our framework proposes an adaptive cognitive load-balancing model that dynamically adjusts task complexity, presentation, and pace based on real-time operator states~\cite{wickens2017mental}. By aligning task demands with operator skills, the proposed approach considers individual strengths, weaknesses, and preferences, thereby allowing for a gradual increase in challenges to promote skill development. For example, during joint reconnaissance missions, tasks requiring high situational awareness can be assigned to experienced operators, and routine data entry is managed by AI. This adaptive approach ensures that operators are not overwhelmed and facilitates effective decision-making in tactical environments.

\subsection{Situational Awareness}

\noindent \textbf{AI-Enhanced Perception}. Traditional situational awareness models in HAT systems often rely on limited or isolated data streams, which results in incomplete or biased situational models. Our framework attempts to overcome these limitations by implementing AI-enhanced perception that integrates diverse data sources, such as radar, satellite imagery, and ground sensors, to create a more holistic and accurate situational model~\cite{endsley2000situation}. The proposed approach leverages multi-sensor fusion techniques, including Bayesian inference, convolutional neural networks for image analysis, and Kalman filtering for real-time sensor data processing, to combine data into coherent and interpretable representations. By merging various sensory inputs, the model maintains consistency and reliability in situational awareness, facilitating seamless communication between AI and human operators. To further enhance real-time situational awareness, the framework incorporates advanced analytics tools that continuously update the situational model, identify emerging threats, and provide timely alerts to operators. This real-time processing capability ensures that sudden changes, such as enemy movements or environmental shifts, are detected and communicated clearly and promptly. In addition, AI-enhanced perception can identify and highlight anomalous patterns that human operators might otherwise overlook, particularly in dynamic tactical environments. For example, in an urban combat simulation, this integration allows AI to identify potential threats overlooked by human operators, thereby offering recommendations that improve decision accuracy and reduce response times. By maintaining an up-to-date, comprehensive situational awareness, the proposed framework can improve decision-making accuracy and enhance the operational readiness of HAT systems.

\vspace{1.0ex}

\noindent \textbf{Shared Mental Models}. In tactical operations, achieving a shared mental model between human operators and AI is important~\cite{stout2017role}. For example, during a coordinated air-ground mission, an AI system can continuously update its situational assessment and communicate key changes to the human team members, ensuring that both AI and humans have an aligned understanding of the mission's progress. Our framework proposes a concept for dynamic shared cognition that can align AI and human understanding in tactical situations to enhance team cohesion and decision-making. This theoretical approach introduces potential mechanisms for comparing and assessing differences between AI and human situational assessments. The concept further outlines possibilities for cognitive synchronization through targeted information exchange to address potential discrepancies in situation understanding.

\subsection{Ethical Considerations}

\noindent \textbf{Ethical Decision Support}. Ethical decision-making in HAT systems typically involves predefined rules or rigid moral frameworks, which limit their adaptability to dynamic tactical situations. In tactical situations, AI-driven HAT can offer proactive ethical decision support by helping human operators navigate legal, moral, and operational implications. For example, during a peacekeeping mission simulation, AI could propose nonlethal measures to de-escalate a conflict while also presenting operators with clear explanations of the ethical implications of each choice. This ensures ethical compliance while enabling informed decision-making. The proposed framework introduces adaptive ethical decision models that integrate ethical guidelines with real-time decision-making processes. These models assess the ethical implications of potential actions and provide human operators with options aligned with international laws and moral standards. Future research should therefore focus on embedding real-time ethical reasoning algorithms that ensure compliance with international laws and offer moral decision support, particularly during conflict situations.

\vspace{1.0ex}

\noindent \textbf{Accountability Frameworks}. The proposed accountability framework clearly defines distributed responsibilities between AI and human team members. This theoretical model introduces possibilities for analyzing team decisions and actions after they occur, allowing for thorough review and assessment. This concept presents potential approaches for tracking and documenting the contributions of AI and human team members to significant operational decisions.

\section{Practical Recommendations}
\label{practical_recommendations}

In addition to proposing a comprehensive framework (Section~\ref{proposed_framework}), this study provides practical recommendations that can contribute to the implementation of AI-driven HAT systems in tactical environments. After exploring the opportunities and challenges of AI-driven HAT for tactical autonomy (Section~\ref{opportunities_and_challenges}), this section provides clear guidance for policymakers, practitioners, and researchers regarding the complexities of HAT development and implementation.

\vspace{1.0ex}
\noindent \textbf{Develop Transparent and \ac{XAI} Systems}. Developing transparent and explainable AI systems is important for building trust between human operators and AI teammates. Interpretable machine learning algorithms can improve the explainability of AI systems~\cite{arrieta2020explainable}. Implementing \ac{XAI} techniques such as generating visualizations or context-based explanations to make AI decision-making processes more transparent, is an example of an actionable step. This empowers human operators to understand the rationale behind AI recommendations and promotes informed collaborative decision-making~\cite{preece2019explainable}. We recommend implementing XAI models that offer visual, auditory, and textual explanations of AI decisions. These models should be tailored to the tactical environment, such as overlaying battlefield maps with AI-generated insights to assist operators in decision-making.

\vspace{1.0ex}
\noindent \textbf{Building Robust AI for Uncertainty}. Given the dynamic and unpredictable nature of tactical environments, \ac{AI} systems should be trained on diverse real-world datasets reflecting various scenarios~\cite{bengio2012practical}. This training enhances the system's ability to handle uncertainty and make context-appropriate decisions. Prioritizing robustness against uncertainty and ambiguity ensures that \ac{AI} systems can effectively adapt to dynamic operational settings. This enables AI systems to make context-appropriate decisions even in unforeseen circumstances.

\vspace{1.0ex}
\noindent \textbf{Human-AI Collaboration}. Effective Human-AI collaboration in tactical autonomy requires a symbiotic relationship between human decision-making and AI assistance. Instead of deploying fully autonomous systems, AI tools should be designed to provide recommendations and options that support human operators, enabling them to make informed decisions while leveraging AI's analytical capabilities. To achieve this, it is important to emphasize human-centric design throughout the development of HAT systems. This approach involves an iterative design process that actively incorporates user feedback at every stage, from conceptual prototyping to field testing. By engaging operators in this process, developers can ensure that AI systems closely align with user needs, thereby enhancing both usability and operator trust. For example, user-centered design frameworks, such as participatory design and co-design workshops, can be employed to gather input directly from operators, allowing AI systems to be tailored to specific operational contexts and user requirements. Training programs for operators are also critical because they help users understand AI's capabilities and limitations, providing more effective collaboration in tactical environments. The iterative development process should include usability testing, interface adjustments, and continuous model training based on real-world interactions. By focusing on human-centric design and continuous user engagement, AI systems can become more intuitive, reduce cognitive load, and facilitate smooth decision-making processes, thereby improving the overall effectiveness of HAT operations.

\vspace{1.0ex}
\noindent \textbf{Emphasize Human-AI Training and Education}. Policymakers and practitioners must invest in developing a comprehensive training program to familiarize human operators with \ac{AI} functionalities and limitations. Designing training programs that cover the full potential of \ac{AI} capabilities, limitations, and best practices for collaboration in tactical operations is important. Using simulation-based training to allow operators to practice in a safe and controlled environment and understand best practices for collaborating with \ac{AI} systems during tactical operations can be effective.

\vspace{1.0ex}
\noindent \textbf{Cybersecurity and Privacy}. Robust cybersecurity measures and data privacy considerations are essential for secure and ethical HAT operations. Policymakers must address cybersecurity concerns by implementing end-to-end encryption for communication among AI-assisted tactical teams. This protects mission-critical information from potential adversaries. In addition, policymakers should examine existing legal and regulatory frameworks, such as the \ac{GDPR}, governing HAT to ensure that they address data privacy concerns~\cite{goodman2017european}.

\vspace{1.0ex}
\noindent \textbf{Continuous Trust Assessment Mechanisms}. Implementing continuous trust assessment mechanisms is important for maintaining operator confidence in AI systems and ensuring effective collaboration. To achieve this, user feedback should be systematically integrated into AI systems by creating feedback loops that allow operators to provide real-time evaluations of AI decisions. For example, real-time feedback interfaces can enable operators to rate AI decisions on reliability, which can be used to dynamically adjust trust levels based on user responses~\cite{lee2004trust}. Trust calibration algorithms also play an important role in evaluating and adjusting trust levels. These algorithms can monitor operator behavior and performance metrics to adjust AI autonomy levels in real-time~\cite{parasuraman2010complacency}. Additionally, physiological monitoring tools, such as sensors for heart rate variability, can be used to assess trust indirectly during interactions with AI systems~\cite{wickens2017mental}. In addition, existing trust measurement frameworks, as discussed in~\cite{jian2000foundations}, offer methods to quantitatively assess trust in automated systems during HAT operations. These frameworks provide empirical insights into operator trust levels by generating real-time reports that help inform adjustments to AI behaviors, thereby fostering more effective human-AI collaboration.

\vspace{1.0ex}

\noindent \textbf{Implement Protocols for System Failures and Recovery}. Designing \ac{AI} systems with fail-safe mechanisms and establishing protocols that allow human operators to regain control in the event of system failures is critical. Human operators must intervene and take control when necessary to ensure operational continuity and safety in dynamic environments.

\vspace{1.0ex}
\noindent \textbf{AI Training for Real-World Scenarios}. Training AI systems using real-world data from tactical operations is important for ensuring accurate, context-aware decision-making. In particular, AI systems should be trained using datasets that simulate realistic operational contexts. For example, using datasets like the DARPA's OFFensive Swarm-Enabled Tactics (OFFSET)~\cite{chung2017offensive} or the Military Operations on Urban Terrain (MOUT)~\cite{scribner2007development} simulation data can significantly enhance AI adaptability in combat situations by replicating battlefield dynamics. Similarly, datasets such as the FEMA dataset\footnote{\url{https://www.fema.gov/openfema-data-page/disaster-declarations-summaries-v2}} on disaster response can prepare AI systems for handling natural or human-made disaster scenarios, improving their capacity for context-specific decision-making. To optimize learning, AI models should incorporate advanced techniques such as reinforcement learning and domain adaptation. Reinforcement learning allows AI systems to interact with simulated environments, enabling them to adapt behaviors based on trial-and-error feedback and learn optimal responses~\cite{bengio2012practical}. Furthermore, domain adaptation can help AI systems generalize from training data to real-world deployment by learning from pre-existing datasets and adapting to new operational environments~\cite{farahani2021brief}. This approach ensures that AI systems can effectively manage unexpected variables commonly encountered in tactical settings.

\vspace{1.0ex}
\noindent \textbf{Develop Ethics and Regulations}. Policymakers and AI researchers must collaborate to establish ethical guidelines and legal frameworks for the use of AI in tactical autonomy. These guidelines should address transparency, accountability, AI bias, and the rights of human operators within human-autonomy teams to ensure the safe, ethical, and responsible deployment of autonomous systems. To achieve this, ethical decision support should be seamlessly integrated into HAT systems, providing operators with real-time ethical guidance and decision options based on international laws and mission rules, even in rapidly evolving scenarios. Ethical decision support modules should be designed to help human operators make legally compliant and morally sound decisions during tactical operations~\cite{arkin2009governing}. These modules can leverage decision trees or rule-based AI systems that embed ethical principles and offer real-time recommendations aligned with international humanitarian laws, such as promoting non-lethal engagement in conflict zones. Drawing from frameworks like Asimov's laws of robotics~\cite{anderson2008asimov, murphy2009beyond}, extended to incorporate modern ethical guidelines for military and emergency operations, these modules present clear decision options that align with mission objectives while adhering to established ethical norms. To further ensure ethical compliance, HAT systems should include compliance-checking mechanisms that continuously monitor AI decisions against legal standards~\cite{anderson2011machine}. Real-time compliance checks can flag actions that may violate ethical guidelines, prompting human operators to review and, if necessary, override AI-generated decisions. These compliance checks can be implemented using rule-based AI systems that reference legal databases in real-time, ensuring that all AI decisions undergo thorough verification before execution. Integrating ethical decision support with compliance monitoring promotes informed decision-making and ensures awareness of legal or ethical risks, thus enabling safe and responsible AI deployment in high-stakes environments~\cite{anderson2011machine}.

\vspace{1.0ex}
\noindent \textbf{Pilot Programs and Cross-Disciplinary Collaborations}. We recommend launching pilot programs that involve a collaborative effort between military agencies, emergency response teams, academic research centers, and technology companies. These pilot programs should focus on testing and refining the proposed AI-driven HAT systems in controlled yet realistic environments. Effective implementation requires cross-disciplinary teams that include AI researchers, human factor experts, ethicists, legal professionals, and tactical operators. By bringing together these diverse perspectives, pilot programs can address the complex challenges of HAT development, such as ethical decision-making, operator trust, and system reliability, thus ensuring that theoretical frameworks are practically applicable. Pilot sites and testing environments should be carefully identified to enhance the impact of these collaborations. Military training centers and disaster response simulation facilities offer controlled settings that replicate real-world scenarios, providing researchers with safe spaces to validate AI performance and identify areas for refinement before broader deployment. Developing clear evaluation criteria and metrics is important for assessing the effectiveness of AI-driven HAT systems during pilot tests. Metrics such as task completion rates, human-operator feedback, and compliance with legal standards can help determine system readiness for real-world applications. By adopting an iterative approach, the collaborations can ensure that the models are not only theoretically sound but also adaptable to real-world conditions, allowing continuous refinement based on pilot outcomes. This approach bridges the gap between theoretical frameworks and practical implementations and contributes to the development of reliable, ethical, and effective HAT systems that satisfy the complex demands of tactical operations.

\section{Conclusion}
\label{conclusion}

This paper has explored the realm of AI-driven human-autonomy collaboration within tactical operations, demonstrating how this integration represents a paradigm shift in decision-making, situational awareness, and operational efficiency. Through our proposed framework, we have provided a structured approach to understanding and advancing AI-driven HAT, organizing key components into four critical areas: trust and transparency, function allocation, situational awareness, and ethical considerations. This framework serves as a foundation for future research and development, offering a systematic way to address the complex challenges of integrating AI into tactical operations. Our exploration of the opportunities and challenges in this domain highlights the transformative potential of AI-driven HAT across diverse sectors, including military operations, emergency response, and law enforcement. The integration of AI technologies offers significant advantages while demanding careful consideration of critical factors, such as trust, transparency, and cognitive load management. Looking forward, it is important to chart a path that embraces the ethical deployment of AI, establishes robust mechanisms for human oversight, and promotes interdisciplinary collaboration among AI researchers, human factor experts, and domain specialists. By addressing these critical challenges and leveraging these opportunities within the proposed framework, we envision a future where AI-powered HAT systems seamlessly integrate human intuition, ethical reasoning, and autonomous capabilities to achieve unprecedented levels of effectiveness in complex tactical environments.

This paper emphasizes the ongoing need for research and development efforts that prioritize human-centric design, transparency, and the establishment of a foundation where AI can serve as an empowering force in tactical autonomy. The proposed framework provides a structured approach for addressing these needs and guiding future developments. In our future work, we aim to investigate approaches for developing and validating scalable HAT systems that address the primary research challenges and knowledge gaps identified in this paper. The proposed system leverages state-of-the-art AI techniques to facilitate seamless collaboration, communication, and coordination between human operators and autonomous systems, emphasizing trust-building, explainability, and cognitive load management.

\section*{Acknowledgment}

This work was supported by the United States DoD Center of Excellence in AI/ML at Howard University under Contract number W911NF-20-2-0277 with the U.S. Army Research Laboratory (ARL). However, any opinions, findings, conclusions, or recommendations expressed in this document are those of the authors and should not be interpreted as representing the official policies, either expressed or implied, of the funding agencies.

\bibliography{HAT_Paper}\addcontentsline{toc}{section}{\refname}

\begin{thebibliography}{100}
\providecommand{\url}[1]{#1}
\csname url@samestyle\endcsname
\providecommand{\newblock}{\relax}
\providecommand{\bibinfo}[2]{#2}
\providecommand{\BIBentrySTDinterwordspacing}{\spaceskip=0pt\relax}
\providecommand{\BIBentryALTinterwordstretchfactor}{4}
\providecommand{\BIBentryALTinterwordspacing}{\spaceskip=\fontdimen2\font plus
\BIBentryALTinterwordstretchfactor\fontdimen3\font minus \fontdimen4\font\relax}
\providecommand{\BIBforeignlanguage}[2]{{%
\expandafter\ifx\csname l@#1\endcsname\relax
\typeout{** WARNING: IEEEtran.bst: No hyphenation pattern has been}%
\typeout{** loaded for the language `#1'. Using the pattern for}%
\typeout{** the default language instead.}%
\else
\language=\csname l@#1\endcsname
\fi
#2}}
\providecommand{\BIBdecl}{\relax}
\BIBdecl

\bibitem{o2023human}
T.~A. O'Neill, C.~Flathmann \emph{et~al.}, ``{Human-autonomy Teaming: Need for a guiding team-based framework?}'' \emph{Computers in Human Behavior}, vol. 146, p. 107762, 2023.

\bibitem{o2022human}
T.~O’Neill, N.~McNeese, A.~Barron, and B.~Schelble, ``{Human--autonomy teaming: A review and analysis of the empirical literature},'' \emph{Human factors}, vol.~64, no.~5, pp. 904--938, 2022.

\bibitem{lyons2021human}
J.~B. Lyons, K.~Sycara, M.~Lewis, and A.~Capiola, ``{Human--autonomy teaming: Definitions, debates, and directions},'' \emph{Frontiers in Psychology}, vol.~12, p. 589585, 2021.

\bibitem{johnson2012autonomy}
M.~Johnson, J.~M. Bradshaw, P.~Feltovich, C.~Jonker, B.~Van~Riemsdijk, and M.~Sierhuis, ``Autonomy and interdependence in human-agent-robot teams,'' \emph{IEEE Intelligent Systems}, vol.~27, no.~2, pp. 43--51, 2012.

\bibitem{wynne2018integrative}
K.~T. Wynne and J.~B. Lyons, ``An integrative model of autonomous agent teammate-likeness,'' \emph{Theoretical Issues in Ergonomics Science}, vol.~19, no.~3, pp. 353--374, 2018.

\bibitem{chen2016human}
J.~Y. Chen, M.~J. Barnes, A.~R. Selkowitz, K.~Stowers, S.~G. Lakhmani, and N.~Kasdaglis, ``Human-autonomy teaming and agent transparency,'' in \emph{Companion Publication of the 21st International Conference on Intelligent User Interfaces}, 2016, pp. 28--31.

\bibitem{endsley2017here}
M.~R. Endsley, ``From here to autonomy: lessons learned from human--automation research,'' \emph{Human factors}, vol.~59, no.~1, pp. 5--27, 2017.

\bibitem{mercado2016intelligent}
J.~E. Mercado, M.~A. Rupp, J.~Y. Chen, M.~J. Barnes, D.~Barber, and K.~Procci, ``{Intelligent agent transparency in human--agent teaming for Multi-UxV management},'' \emph{Human factors}, vol.~58, no.~3, pp. 401--415, 2016.

\bibitem{myers2018autonomous}
C.~Myers, J.~Ball, N.~Cooke, M.~Freiman, M.~Caisse, S.~Rodgers, M.~Demir, and N.~McNeese, ``Autonomous intelligent agents for team training,'' \emph{IEEE Intelligent Systems}, vol.~34, no.~2, pp. 3--14, 2018.

\bibitem{klare2019autonomous}
M.~T. Klare, ``Autonomous weapons systems and the laws of war,'' \emph{Arms Control Today}, vol.~49, no.~2, pp. 6--12, 2019.

\bibitem{watson2005autonomous}
D.~P. Watson and D.~H. Scheidt, ``Autonomous systems,'' \emph{Johns Hopkins APL technical digest}, vol.~26, no.~4, pp. 368--376, 2005.

\bibitem{hagos2022recent}
D.~H. Hagos and D.~B. Rawat, ``{Recent Advances in Artificial Intelligence and Tactical Autonomy: Current Status, Challenges, and Perspectives},'' \emph{Sensors}, vol.~22, no.~24, p. 9916, 2022.

\bibitem{shively2022human}
J.~Shively, ``{Human Autonomy Teaming: Principles and Applications},'' in \emph{5th Annual Human Systems Integration (HSI) Summit: Human Machine Interfaces}, 2022.

\bibitem{beaucillon2023special}
C.~Beaucillon and S.~Poli, ``{Special Focus on EU Strategic Autonomy and Technological Sovereignty: An Introduction},'' \emph{European Papers-A Journal on Law and Integration}, vol. 2023, no.~2, pp. 411--416, 2023.

\bibitem{united1994department}
U.~S. J.~C. of~Staff, \emph{{Department of Defense Dictionary of Military and Associated Terms}}.\hskip 1em plus 0.5em minus 0.4em\relax Joint Chiefs of Staff, 1994, vol.~1, no.~2.

\bibitem{demir2018team}
M.~Demir, A.~D. Likens, N.~J. Cooke, P.~G. Amazeen, and N.~J. McNeese, ``Team coordination and effectiveness in human-autonomy teaming,'' \emph{IEEE Transactions on Human-Machine Systems}, vol.~49, no.~2, pp. 150--159, 2018.

\bibitem{shively2018human}
R.~J. Shively, J.~Lachter, S.~L. Brandt, M.~Matessa, V.~Battiste, and W.~W. Johnson, ``Why human-autonomy teaming?'' in \emph{Advances in Neuroergonomics and Cognitive Engineering: Proceedings of the AHFE 2017 International Conference on Neuroergonomics and Cognitive Engineering, July 17--21, 2017, The Westin Bonaventure Hotel, Los Angeles, California, USA 8}.\hskip 1em plus 0.5em minus 0.4em\relax Springer, 2018, pp. 3--11.

\bibitem{roth2019function}
E.~M. Roth, C.~Sushereba, L.~G. Militello, J.~Diiulio, and K.~Ernst, ``Function allocation considerations in the era of human autonomy teaming,'' \emph{Journal of Cognitive Engineering and Decision Making}, vol.~13, no.~4, pp. 199--220, 2019.

\bibitem{schelble2020towards}
B.~G. Schelble, C.~Flathmann, and N.~McNeese, ``Towards meaningfully integrating human-autonomy teaming in applied settings,'' in \emph{Proceedings of the 8th international conference on human-agent interaction}, 2020, pp. 149--156.

\bibitem{lucero2020human}
C.~Lucero, C.~Izumigawa, K.~Frederiksen, L.~Nans, R.~Iden, and D.~S. Lange, ``{Human-autonomy teaming and explainable AI capabilities in rts games},'' in \emph{Engineering Psychology and Cognitive Ergonomics. Cognition and Design: 17th International Conference, EPCE 2020, Held as Part of the 22nd HCI International Conference, HCII 2020, Copenhagen, Denmark, July 19--24, 2020, Proceedings, Part II 22}.\hskip 1em plus 0.5em minus 0.4em\relax Springer, 2020, pp. 161--171.

\bibitem{teaming2022state}
\BIBentryALTinterwordspacing
N.~A. of~Sciences, ``{Human-AI Teaming: State-of-the-Art and Research Needs},'' National Academies of Sciences, Engineering, and Medicine, Washington, D.C., 2022. [Online]. Available: \url{https://doi.org/10.17226/26355}
\BIBentrySTDinterwordspacing

\bibitem{veitch2022systematic}
E.~Veitch and O.~A. Alsos, ``{A systematic review of human-AI interaction in autonomous ship systems},'' \emph{Safety science}, vol. 152, p. 105778, 2022.

\bibitem{ferguson2003autonomous}
D.~Ferguson, A.~Morris, D.~Haehnel, C.~Baker, Z.~Omohundro, C.~Reverte, S.~Thayer, C.~Whittaker, W.~Whittaker, W.~Burgard \emph{et~al.}, ``An autonomous robotic system for mapping abandoned mines,'' \emph{Advances in Neural Information Processing Systems}, vol.~16, 2003.

\bibitem{schaefer2021human}
K.~E. Schaefer, B.~Perelman, J.~Rexwinkle, J.~Canady, C.~Neubauer, N.~Waytowich, G.~Larkin, K.~Cox, M.~Geuss, G.~Gremillion \emph{et~al.}, ``{Human-autonomy teaming for the tactical edge: The importance of humans in artificial intelligence research and development},'' in \emph{Systems Engineering and Artificial Intelligence}.\hskip 1em plus 0.5em minus 0.4em\relax Springer, 2021, pp. 115--148.

\bibitem{van2018human}
K.~Van Den~Bosch, A.~Bronkhorst \emph{et~al.}, ``{Human-AI cooperation to benefit military decision making}.''\hskip 1em plus 0.5em minus 0.4em\relax NATO, 2018.

\bibitem{schubert2018artificial}
J.~Schubert, J.~Brynielsson, M.~Nilsson, and P.~Svenmarck, ``Artificial intelligence for decision support in command and control systems,'' in \emph{Proceedings of the 23rd International Command and Control Research \& Technology Symposium Multi-Domain C2}, 2018, pp. 18--33.

\bibitem{funke2022teamwork}
G.~Funke, J.~B. Lyons, E.~T. Greenlee, M.~T. Tolston, and G.~Matthews, ``Teamwork in human-machine teaming,'' \emph{Frontiers in Psychology}, vol.~13, p. 999000, 2022.

\bibitem{mctear2022conversational}
M.~McTear, \emph{{Conversational AI: Dialogue systems, conversational agents, and chatbots}}.\hskip 1em plus 0.5em minus 0.4em\relax Springer Nature, 2022.

\bibitem{meszaros2018trusted}
E.~L. Meszaros, L.~R. Le~Vie, and B.~D. Allen, ``{Trusted Communication: Utilizing Speech Communication to Enhance Human-Machine Teaming Success},'' in \emph{2018 Aviation Technology, Integration, and Operations Conference}, 2018, p. 4014.

\bibitem{peeters2021hybrid}
M.~M. Peeters, J.~van Diggelen, K.~Van Den~Bosch, A.~Bronkhorst, M.~A. Neerincx, J.~M. Schraagen, and S.~Raaijmakers, ``{Hybrid collective intelligence in a human--AI society},'' \emph{AI \& society}, vol.~36, pp. 217--238, 2021.

\bibitem{van2021delegation}
J.~van Diggelen, J.~Barnhoorn, R.~Post, J.~Sijs, N.~van~der Stap, and J.~van~der Waa, ``Delegation in human-machine teaming: progress, challenges and prospects,'' in \emph{Intelligent Human Systems Integration 2021: Proceedings of the 4th International Conference on Intelligent Human Systems Integration (IHSI 2021): Integrating People and Intelligent Systems, February 22-24, 2021, Palermo, Italy}.\hskip 1em plus 0.5em minus 0.4em\relax Springer, 2021, pp. 10--16.

\bibitem{castelfranchi1998towards}
C.~Castelfranchi and R.~Falcone, ``Towards a theory of delegation for agent-based systems,'' \emph{Robotics and Autonomous systems}, vol.~24, no. 3-4, pp. 141--157, 1998.

\bibitem{cummings2011impact}
M.~L. Cummings, J.~P. How, A.~Whitten, and O.~Toupet, ``The impact of human--automation collaboration in decentralized multiple unmanned vehicle control,'' \emph{Proc. of the IEEE}, vol. 100, no.~3, pp. 660--671, 2011.

\bibitem{alotaibi2019lsar}
E.~T. Alotaibi, S.~S. Alqefari, and A.~Koubaa, ``Lsar: Multi-uav collaboration for search and rescue missions,'' \emph{IEEE Access}, vol.~7, pp. 55\,817--55\,832, 2019.

\bibitem{scherer2015autonomous}
J.~Scherer, S.~Yahyanejad, S.~Hayat, E.~Yanmaz, T.~Andre, A.~Khan, V.~Vukadinovic, C.~Bettstetter, H.~Hellwagner, and B.~Rinner, ``{An autonomous multi-UAV system for search and rescue},'' in \emph{Proceedings of the first workshop on micro aerial vehicle networks, systems, and applications for civilian use}, 2015, pp. 33--38.

\bibitem{shakhatreh2019unmanned}
H.~Shakhatreh, A.~H. Sawalmeh, A.~Al-Fuqaha, Z.~Dou, E.~Almaita, I.~Khalil, N.~S. Othman, A.~Khreishah, and M.~Guizani, ``{Unmanned aerial vehicles (UAVs): A survey on civil applications and key research challenges},'' \emph{Ieee Access}, vol.~7, pp. 48\,572--48\,634, 2019.

\bibitem{lattanzi2017review}
D.~Lattanzi and G.~Miller, ``Review of robotic infrastructure inspection systems,'' \emph{Journal of Infrastructure Systems}, vol.~23, no.~3, 2017.

\bibitem{menouar2017uav}
H.~Menouar, I.~Guvenc, K.~Akkaya, A.~S. Uluagac, A.~Kadri, and A.~Tuncer, ``{UAV-enabled intelligent transportation systems for the smart city: Applications and challenges},'' \emph{IEEE Communications Magazine}, vol.~55, no.~3, pp. 22--28, 2017.

\bibitem{huang2013development}
Y.~Huang, S.~J. Thomson, W.~C. Hoffmann, Y.~Lan, and B.~K. Fritz, ``Development and prospect of unmanned aerial vehicle technologies for agricultural production management,'' \emph{International Journal of Agricultural and Biological Engineering}, vol.~6, no.~3, pp. 1--10, 2013.

\bibitem{pearce2018optimizing}
M.~Pearce, B.~Mutlu, J.~Shah, and R.~Radwin, ``Optimizing makespan and ergonomics in integrating collaborative robots into manufacturing processes,'' \emph{IEEE transactions on automation science and engineering}, vol.~15, no.~4, pp. 1772--1784, 2018.

\bibitem{li2023proactive}
S.~Li, P.~Zheng, S.~Liu, Z.~Wang, X.~V. Wang, L.~Zheng, and L.~Wang, ``{Proactive human--robot collaboration: Mutual-cognitive, predictable, and self-organising perspectives},'' \emph{Robotics and Computer-Integrated Manufacturing}, vol.~81, p. 102510, 2023.

\bibitem{okamura2010medical}
A.~M. Okamura, M.~J. Matari{\'c}, and H.~I. Christensen, ``Medical and health-care robotics,'' \emph{IEEE Robotics \& Automation Magazine}, vol.~17, no.~3, pp. 26--37, 2010.

\bibitem{lei2022human}
T.~Lei, P.~Chintam, D.~W. Carruth, G.~E. Jan, and C.~Luo, ``Human-autonomy teaming-based robot informative path planning and mapping algorithms with tree search mechanism,'' in \emph{IEEE 3rd Int. Conference on Human-Machine Systems}.\hskip 1em plus 0.5em minus 0.4em\relax IEEE, 2022, pp. 1--6.

\bibitem{hussain2014use}
A.~Hussain, A.~Malik, M.~U. Halim, and A.~M. Ali, ``The use of robotics in surgery: a review,'' \emph{International journal of clinical practice}, vol.~68, no.~11, pp. 1376--1382, 2014.

\bibitem{howe1999robotics}
R.~D. Howe and Y.~Matsuoka, ``Robotics for surgery,'' \emph{Annual review of biomedical engineering}, vol.~1, no.~1, pp. 211--240, 1999.

\bibitem{davies2000review}
B.~Davies, ``A review of robotics in surgery,'' \emph{Proceedings of the Institution of Mechanical Engineers, Part H: Journal of Engineering in Medicine}, vol. 214, no.~1, pp. 129--140, 2000.

\bibitem{wang2020decision}
W.~Wang, X.~Na, D.~Cao, J.~Gong, J.~Xi, Y.~Xing, and F.-Y. Wang, ``{Decision-making in driver-automation shared control: A review and perspectives},'' \emph{IEEE/CAA Journal of Automatica Sinica}, vol.~7, no.~5, pp. 1289--1307, 2020.

\bibitem{blaschke2009driver}
C.~Blaschke, F.~Breyer, B.~F{\"a}rber, J.~Freyer, and R.~Limbacher, ``Driver distraction based lane-keeping assistance,'' \emph{Transportation research part F: traffic psychology and behaviour}, vol.~12, no.~4, pp. 288--299, 2009.

\bibitem{saito2016driver}
Y.~Saito, M.~Itoh, and T.~Inagaki, ``Driver assistance system with a dual control scheme: Effectiveness of identifying driver drowsiness and preventing lane departure accidents,'' \emph{IEEE Transactions on Human-Machine Systems}, vol.~46, no.~5, pp. 660--671, 2016.

\bibitem{tada2016simultaneous}
S.~Tada, K.~Sonoda, and T.~Wada, ``Simultaneous achievement of workload reduction and skill enhancement in backward parking by haptic guidance,'' \emph{IEEE Transactions on Intelligent Vehicles}, vol.~1, no.~4, pp. 292--301, 2016.

\bibitem{wang2016human}
W.~Wang, J.~Xi, C.~Liu, and X.~Li, ``Human-centered feed-forward control of a vehicle steering system based on a driver's path-following characteristics,'' \emph{IEEE transactions on intelligent transportation systems}, vol.~18, no.~6, pp. 1440--1453, 2016.

\bibitem{wang2015human}
W.~Wang, J.~Xi, and J.~Wang, ``Human-centered feed-forward control of a vehicle steering system based on a driver's steering model,'' in \emph{2015 American Control Conference (ACC)}.\hskip 1em plus 0.5em minus 0.4em\relax IEEE, 2015, pp. 3361--3366.

\bibitem{ercan2018predictive}
Z.~Ercan, A.~Carvalho, H.~E. Tseng, M.~G{\"o}ka{\c{s}}an, and F.~Borrelli, ``A predictive control framework for torque-based steering assistance to improve safety in highway driving,'' \emph{Vehicle system dynamics}, vol.~56, no.~5, pp. 810--831, 2018.

\bibitem{sarter2023contributions}
N.~Sarter, K.~Panesar, A.~Bhardwaj \emph{et~al.}, ``{The Contributions of Human Operators to Safety and Risk Mitigation: Implications for Crew Complements and Automation/Autonomy Levels in Commercial Transport Operations},'' United States. Department of Transportation. Federal Aviation Administration, Tech. Rep., 2023.

\bibitem{tokadli2022autonomy}
G.~Tokadl{\i} and M.~C. Dorneich, ``{Autonomy as a teammate: Evaluation of teammate-likeness},'' \emph{Journal of Cognitive Engineering and Decision Making}, vol.~16, no.~4, pp. 282--300, 2022.

\bibitem{el2023joint}
H.~El~Alami, M.~Nwosu, and D.~B. Rawat, ``Joint human and autonomy teaming for defense: status, challenges, and perspectives,'' \emph{Artificial Intelligence and Machine Learning for Multi-Domain Operations Applications V}, vol. 12538, pp. 144--158, 2023.

\bibitem{schelble2023investigating}
B.~G. Schelble, C.~Flathmann, N.~J. McNeese, T.~O’Neill, R.~Pak, and M.~Namara, ``Investigating the effects of perceived teammate artificiality on human performance and cognition,'' \emph{International Journal of Human--Computer Interaction}, vol.~39, no.~13, pp. 2686--2701, 2023.

\bibitem{lee2004trust}
J.~D. Lee and K.~A. See, ``{Trust in automation: Designing for appropriate reliance},'' \emph{Human factors}, vol.~46, no.~1, pp. 50--80, 2004.

\bibitem{chen2018human}
J.~Y. Chen, ``Human-autonomy teaming in military settings,'' \emph{Theoretical issues in ergonomics science}, vol.~19, no.~3, 2018.

\bibitem{hughes2016saving}
A.~M. Hughes, M.~E. Gregory, D.~L. Joseph, S.~C. Sonesh, S.~L. Marlow, C.~N. Lacerenza, L.~E. Benishek, H.~B. King, and E.~Salas, ``{Saving lives: A meta-analysis of team training in healthcare.}'' \emph{Journal of applied psychology}, vol. 101, no.~9, p. 1266, 2016.

\bibitem{izumigawa2020building}
C.~Izumigawa, C.~Lucero, L.~Nans, K.~Frederiksen, O.~Hui, I.~Enriquez, S.~Rothman, and R.~Iden, ``{Building Human-Autonomy Teaming Aids for Real-Time Strategy Games},'' in \emph{HCI in Games: Second International Conference, HCI-Games 2020, Held as Part of the 22nd HCI International Conference, HCII 2020, Copenhagen, Denmark, July 19--24, 2020, Proceedings 22}.\hskip 1em plus 0.5em minus 0.4em\relax Springer, 2020, pp. 117--127.

\bibitem{strybel2018effectiveness}
T.~Z. Strybel, J.~Keeler, V.~Barakezyan, A.~Alvarez, N.~Mattoon, K.-P.~L. Vu, and V.~Battiste, ``Effectiveness of human autonomy teaming in cockpit applications,'' in \emph{Human Interface and the Management of Information. Information in Applications and Services: 20th International Conference, HIMI 2018, Held as Part of HCI International 2018, Las Vegas, NV, USA, July 15-20, 2018, Proceedings, Part II 20}.\hskip 1em plus 0.5em minus 0.4em\relax Springer, 2018, pp. 465--476.

\bibitem{r2016application}
S.~R.~Jay, S.~L. Brandt, J.~Lachter, M.~Matessa, G.~Sadler, and H.~Battiste, ``{Application of human-autonomy teaming (HAT) patterns to reduced crew operations (RCO)},'' in \emph{Engineering Psychology and Cognitive Ergonomics: 13th International Conference, EPCE 2016, Held as Part of HCI International 2016, Toronto, ON, Canada, July 17-22, 2016, Proceedings 13}.\hskip 1em plus 0.5em minus 0.4em\relax Springer, 2016, pp. 244--255.

\bibitem{goyal2018urban}
R.~Goyal, C.~Reiche, C.~Fernando, J.~Serrao, S.~Kimmel, A.~Cohen, and S.~Shaheen, ``{Urban air mobility (UAM) market study},'' NASA, Tech. Rep., 2018.

\bibitem{pazouki2018investigation}
K.~Pazouki, N.~Forbes, R.~A. Norman, and M.~D. Woodward, ``Investigation on the impact of human-automation interaction in maritime operations,'' \emph{Ocean engineering}, vol. 153, pp. 297--304, 2018.

\bibitem{parasuraman2017humans}
R.~Parasuraman and C.~D. Wickens, ``{Humans: Still vital after all these years of automation},'' in \emph{Decision Making in Aviation}.\hskip 1em plus 0.5em minus 0.4em\relax Routledge, 2017, pp. 251--260.

\bibitem{onken1997cockpit}
R.~Onken, ``{The cockpit assistant system CASSY as an on-board player in the ATM environment},'' in \emph{Proceedings of first air traffic management research and development seminar. Saclay, France}, 1997.

\bibitem{national2021human}
{National Academies of Sciences, Engineering, and Medicine and others}, ``Human-ai teaming: State-of-the-art and research needs,'' 2021.

\bibitem{jones2011using}
R.~E. Jones, E.~S. Connors, M.~E. Mossey, J.~R. Hyatt, N.~J. Hansen, and M.~R. Endsley, ``Using fuzzy cognitive mapping techniques to model situation awareness for army infantry platoon leaders,'' \emph{Computational and Mathematical Organization Theory}, vol.~17, pp. 272--295, 2011.

\bibitem{kokar2012situation}
M.~M. Kokar and M.~R. Endsley, ``Situation awareness and cognitive modeling,'' \emph{IEEE Intelligent Systems}, vol.~27, no.~3, pp. 91--96, 2012.

\bibitem{paleja2021utility}
R.~Paleja, M.~Ghuy, N.~Ranawaka~Arachchige, R.~Jensen, and M.~Gombolay, ``The utility of explainable ai in ad hoc human-machine teaming,'' \emph{Advances in neural information processing systems}, vol.~34, pp. 610--623, 2021.

\bibitem{xu2023transitioning}
W.~Xu, M.~J. Dainoff, L.~Ge, and Z.~Gao, ``{Transitioning to human interaction with AI systems: New challenges and opportunities for HCI professionals to enable human-centered AI},'' \emph{International Journal of Human--Computer Interaction}, vol.~39, no.~3, pp. 494--518, 2023.

\bibitem{carroll2019utility}
M.~Carroll, R.~Shah, M.~K. Ho, T.~Griffiths, S.~Seshia, P.~Abbeel, and A.~Dragan, ``On the utility of learning about humans for human-ai coordination,'' \emph{Advances in neural information processing systems}, vol.~32, 2019.

\bibitem{dierdorff2019power}
E.~C. Dierdorff, D.~M. Fisher, and R.~S. Rubin, ``The power of percipience: Consequences of self-awareness in teams on team-level functioning and performance,'' \emph{Journal of Management}, vol.~45, no.~7, pp. 2891--2919, 2019.

\bibitem{dorneich2017evaluation}
M.~C. Dorneich, B.~Passinger, C.~Hamblin, C.~Keinrath, J.~Va{\v{s}}ek, S.~D. Whitlow, and M.~Beekhuyzen, ``Evaluation of the display of cognitive state feedback to drive adaptive task sharing,'' \emph{Frontiers in neuroscience}, vol.~11, p. 144, 2017.

\bibitem{chella2020developing}
A.~Chella, A.~Pipitone, A.~Morin, and F.~Racy, ``Developing self-awareness in robots via inner speech,'' \emph{Frontiers in Robotics and AI}, vol.~7, p.~16, 2020.

\bibitem{nofi2000defining}
A.~A. Nofi, ``Defining and measuring shared situational awareness,'' 2000.

\bibitem{harrald2007shared}
J.~Harrald and T.~Jefferson, ``Shared situational awareness in emergency management mitigation and response,'' in \emph{2007 40th Annual Hawaii International Conference on System Sciences (HICSS'07)}.\hskip 1em plus 0.5em minus 0.4em\relax IEEE, 2007, pp. 23--23.

\bibitem{salmon2017distributed}
P.~M. Salmon, N.~A. Stanton, and D.~P. Jenkins, ``{Distributed situation awareness: Theory, measurement and application to teamwork},'' 2017.

\bibitem{salas2017situation}
E.~Salas, C.~Prince, D.~P. Baker, and L.~Shrestha, ``{Situation awareness in team performance: Implications for measurement and training},'' \emph{Situational awareness}, pp. 63--76, 2017.

\bibitem{stout2017role}
R.~J. Stout, J.~A. Cannon-Bowers, and E.~Salas, ``The role of shared mental models in developing team situational awareness: Implications for training,'' in \emph{Situational awareness}.\hskip 1em plus 0.5em minus 0.4em\relax Routledge, 2017, pp. 287--318.

\bibitem{yeasmin2019benefits}
S.~Yeasmin, ``Benefits of artificial intelligence in medicine,'' in \emph{2019 2nd International Conference on Computer Applications \& Information Security (ICCAIS)}.\hskip 1em plus 0.5em minus 0.4em\relax IEEE, 2019, pp. 1--6.

\bibitem{mitic2019benefits}
V.~Miti{\'c} \emph{et~al.}, ``Benefits of artificial intelligence and machine learning in marketing,'' in \emph{Sinteza 2019-International scientific conference on information technology and data related research}.\hskip 1em plus 0.5em minus 0.4em\relax Singidunum University, 2019, pp. 472--477.

\bibitem{jain2020transforming}
P.~Jain and K.~Aggarwal, ``Transforming marketing with artificial intelligence,'' \emph{International Research Journal of Engineering and Technology}, vol.~7, no.~7, pp. 3964--3976, 2020.

\bibitem{vorm2020computer}
E.~S. Vorm, ``{Computer-centered humans: why human-AI interaction research will be critical to successful AI integration in the DoD},'' \emph{IEEE Intelligent Systems}, vol.~35, no.~4, pp. 112--116, 2020.

\bibitem{simpson2021real}
T.~Simpson, ``Real-time drone surveillance system for violent crowd behavior unmanned aircraft system (uas)--human autonomy teaming (hat),'' in \emph{2021 IEEE/AIAA 40th Digital Avionics Systems Conference (DASC)}.\hskip 1em plus 0.5em minus 0.4em\relax IEEE, 2021, pp. 1--9.

\bibitem{endsley2023supporting}
M.~R. Endsley, ``{Supporting Human-AI Teams: Transparency, explainability, and situation awareness},'' \emph{Computers in Human Behavior}, vol. 140, p. 107574, 2023.

\bibitem{robbemond2022understanding}
V.~Robbemond, O.~Inel, and U.~Gadiraju, ``{Understanding the Role of Explanation Modality in AI-assisted Decision-making},'' in \emph{Proceedings of the 30th ACM Conference on User Modeling, Adaptation and Personalization}, 2022, pp. 223--233.

\bibitem{samek2019explainable}
W.~Samek, G.~Montavon, A.~Vedaldi, L.~K. Hansen, and K.-R. M{\"u}ller, \emph{{Explainable AI: interpreting, explaining and visualizing deep learning}}.\hskip 1em plus 0.5em minus 0.4em\relax Springer Nature, 2019, vol. 11700.

\bibitem{ehsan2021expanding}
U.~Ehsan, Q.~V. Liao, M.~Muller, M.~O. Riedl, and J.~D. Weisz, ``{Expanding explainability: Towards social transparency in ai systems},'' in \emph{Proceedings of the 2021 CHI Conference on Human Factors in Computing Systems}, 2021, pp. 1--19.

\bibitem{guidotti2018survey}
R.~Guidotti, A.~Monreale, S.~Ruggieri, F.~Turini, F.~Giannotti, and D.~Pedreschi, ``A survey of methods for explaining black box models,'' \emph{ACM computing surveys (CSUR)}, vol.~51, no.~5, pp. 1--42, 2018.

\bibitem{alix2021empowering}
C.~Alix, D.~Lafond, J.~Mattioli, J.~De~Heer, M.~Chattington, and P.-O. Robic, ``Empowering adaptive human autonomy collaboration with artificial intelligence,'' in \emph{2021 16th International Conference of System of Systems Engineering (SoSE)}.\hskip 1em plus 0.5em minus 0.4em\relax IEEE, 2021, pp. 126--131.

\bibitem{dubey2020haco}
A.~Dubey, K.~Abhinav, S.~Jain, V.~Arora, and A.~Puttaveerana, ``{HACO: a framework for developing human-AI teaming},'' in \emph{Proceedings of the 13th Innovations in Software Engineering Conference on Formerly known as India Software Engineering Conference}, 2020, pp. 1--9.

\bibitem{endsley1995toward}
M.~R. Endsley, ``Toward a theory of situation awareness in dynamic systems,'' \emph{Human factors}, vol.~37, no.~1, pp. 32--64, 1995.

\bibitem{lafond2014hci}
D.~Lafond, R.~Proulx, A.~Morris, W.~Ross, A.~Bergeron-Guyard, and M.~Ulieru, ``{HCI dilemmas for context-aware support in intelligence analysis},'' in \emph{Adapt. 2014, Sixth Int. Conf. Adapt. Self-Adaptive Syst. Appl}, 2014, pp. 68--72.

\bibitem{miller2007designing}
C.~A. Miller and R.~Parasuraman, ``{Designing for flexible interaction between humans and automation: Delegation interfaces for supervisory control},'' \emph{Human factors}, vol.~49, no.~1, pp. 57--75, 2007.

\bibitem{erol2016tangible}
S.~Erol, A.~J{\"a}ger, P.~Hold, K.~Ott, and W.~Sihn, ``{Tangible Industry 4.0: a scenario-based approach to learning for the future of production},'' \emph{Procedia CiRp}, vol.~54, pp. 13--18, 2016.

\bibitem{li2023trustworthy}
B.~Li, P.~Qi, B.~Liu, S.~Di, J.~Liu, J.~Pei, J.~Yi, and B.~Zhou, ``{Trustworthy AI: From principles to practices},'' \emph{ACM Computing Surveys}, vol.~55, no.~9, pp. 1--46, 2023.

\bibitem{zaki2021reliability}
O.~Zaki, M.~Dunnigan, V.~Robu, and D.~Flynn, ``Reliability and safety of autonomous systems based on semantic modelling for self-certification,'' \emph{Robotics}, vol.~10, no.~1, p.~10, 2021.

\bibitem{ramos2019autonomous}
M.~A. Ramos, C.~A. Thieme, I.~B. Utne, and A.~Mosleh, ``Autonomous systems safety--state of the art and challenges,'' in \emph{Proceedings of the First Int. Workshop on Autonomous Systems Safety}.\hskip 1em plus 0.5em minus 0.4em\relax NTNU, 2019.

\bibitem{alexander2018state}
R.~D. Alexander, R.~Ashmore, and A.~Banks, ``The state of solutions for autonomous systems safety,'' 2018.

\bibitem{andrews2023role}
R.~W. Andrews, J.~M. Lilly, D.~Srivastava \emph{et~al.}, ``{The role of shared mental models in human-AI teams: a theoretical review},'' \emph{Theoretical Issues in Ergonomics Science}, vol.~24, no.~2, pp. 129--175, 2023.

\bibitem{bansal2019beyond}
G.~Bansal, B.~Nushi, E.~Kamar, W.~S. Lasecki, D.~S. Weld, and E.~Horvitz, ``{Beyond accuracy: The role of mental models in human-AI team performance},'' in \emph{Proceedings of the AAAI conference on human computation and crowdsourcing}, vol.~7, no.~1, 2019, pp. 2--11.

\bibitem{kaur2019building}
H.~Kaur, A.~Williams, and W.~Lasecki, ``Building shared mental models between humans and ai for effective collaboration,'' \emph{CHI'19}, 2019.

\bibitem{schelble2020designing}
B.~Schelble, L.-B. Canonico, N.~McNeese, J.~Carroll, and C.~Hird, ``Designing human-autonomy teaming experiments through reinforcement learning,'' in \emph{Proceedings of the Human Factors and Ergonomics Society Annual Meeting}, vol.~64, no.~1.\hskip 1em plus 0.5em minus 0.4em\relax SAGE Publications Sage CA: Los Angeles, CA, 2020, pp. 1426--1430.

\bibitem{huang2020human}
L.~Huang, N.~Cooke, C.~Johnson, G.~Lematta, S.~Bhatti, M.~Barnes, and E.~Holder, ``{Human-Autonomy Teaming: Interaction Metrics and Models for Next Generation Combat Vehicle Concepts},'' ARIZONA STATE UNIV EAST MESA AZ, Tech. Rep., 2020.

\bibitem{cummings2014man}
M.~M. Cummings, ``Man versus machine or man+ machine?'' \emph{IEEE Intelligent Systems}, vol.~29, no.~5, pp. 62--69, 2014.

\bibitem{metzger2005automation}
U.~Metzger and R.~Parasuraman, ``{Automation in future air traffic management: Effects of decision aid reliability on controller performance and mental workload},'' \emph{Human factors}, vol.~47, no.~1, pp. 35--49, 2005.

\bibitem{richards2020measure}
D.~Richards, ``{Measure for Measure: How do we assess human autonomy teaming?}'' in \emph{HCI International 2020--Late Breaking Papers: Cognition, Learning and Games: 22nd HCI International Conference, HCII 2020, Copenhagen, Denmark, July 19--24, 2020, Proceedings 22}.\hskip 1em plus 0.5em minus 0.4em\relax Springer, 2020, pp. 227--239.

\bibitem{billings1996human}
C.~E. Billings, ``{Human-centered aviation automation: Principles and guidelines},'' NASA, Tech. Rep., 1996.

\bibitem{arkin2009governing}
R.~Arkin, \emph{Governing lethal behavior in autonomous robots}.\hskip 1em plus 0.5em minus 0.4em\relax CRC press, 2009.

\bibitem{boardman2019exploration}
M.~Boardman and F.~Butcher, ``{An exploration of maintaining human control in AI enabled systems and the challenges of achieving it},'' in \emph{Workshop on Big Data Challenge-Situation Awareness and Decision Support. Brussels: North Atlantic Treaty Organization Science and Technology Organization. Porton Down: Dstl Porton Down}, 2019.

\bibitem{mcneese2019understanding}
N.~McNeese, M.~Demir, E.~Chiou, N.~Cooke, and G.~Yanikian, ``Understanding the role of trust in human-autonomy teaming,'' \emph{Proceedings of the 52nd Hawaii International Conference on SystemSciences.}, 2019.

\bibitem{balasubramaniam2023transparency}
N.~Balasubramaniam, M.~Kauppinen, A.~Rannisto, K.~Hiekkanen, and S.~Kujala, ``{Transparency and explainability of AI systems: From ethical guidelines to requirements},'' \emph{Information and Software Technology}, vol. 159, p. 107197, 2023.

\bibitem{larsson2020transparency}
S.~Larsson and F.~Heintz, ``Transparency in artificial intelligence,'' \emph{Internet Policy Review}, vol.~9, no.~2, 2020.

\bibitem{hou2021impacts}
M.~Hou, G.~Ho, and D.~Dunwoody, ``{IMPACTS: A trust model for human-autonomy teaming},'' \emph{Human-Intelligent Systems Integration}, pp. 1--19, 2021.

\bibitem{asatiani2020challenges}
A.~Asatiani, P.~Malo, P.~R. Nagb{\o}l, E.~Penttinen, T.~Rinta-Kahila, and A.~Salovaara, ``{Challenges of explaining the behavior of black-box AI systems},'' \emph{MIS Quarterly Executive}, vol.~19, no.~4, pp. 259--278, 2020.

\bibitem{el2016mining}
I.~El~Moudden, M.~Ouzir, B.~Benyacoub, and S.~ElBernoussi, ``Mining human activity using dimensionality reduction and pattern recognition,'' \emph{Contemporary Engineering Sciences}, vol.~9, no.~21, pp. 1031--1041, 2016.

\bibitem{trunk1979problem}
G.~V. Trunk, ``{A problem of dimensionality: A simple example},'' \emph{IEEE Transactions on pattern analysis and machine intelligence}, no.~3, pp. 306--307, 1979.

\bibitem{olson2013exploration}
E.~Olson, J.~Strom, R.~Goeddel, R.~Morton, P.~Ranganathan, and A.~Richardson, ``Exploration and mapping with autonomous robot teams,'' \emph{Communications of the ACM}, vol.~56, no.~3, pp. 62--70, 2013.

\bibitem{marathe2014confidence}
A.~R. Marathe, B.~J. Lance, K.~McDowell, W.~D. Nothwang, and J.~S. Metcalfe, ``Confidence metrics improve human-autonomy integration,'' in \emph{Proceedings of the 2014 ACM/IEEE international conference on Human-robot interaction}, 2014, pp. 240--241.

\bibitem{eitel2021beyond}
R.~Eitel-Porter, ``{Beyond the promise: implementing ethical AI},'' \emph{AI and Ethics}, vol.~1, pp. 73--80, 2021.

\bibitem{benjamins2021choices}
R.~Benjamins, ``{A choices framework for the responsible use of AI},'' \emph{AI and Ethics}, vol.~1, no.~1, pp. 49--53, 2021.

\bibitem{stanton2006distributed}
N.~A. Stanton, R.~Stewart, D.~Harris, R.~J. Houghton, C.~Baber, R.~McMaster, P.~Salmon, G.~Hoyle, G.~Walker, M.~S. Young \emph{et~al.}, ``Distributed situation awareness in dynamic systems: theoretical development and application of an ergonomics methodology,'' \emph{Ergonomics}, vol.~49, no. 12-13, pp. 1288--1311, 2006.

\bibitem{parasuraman2009adaptive}
R.~Parasuraman, K.~A. Cosenzo, and E.~De~Visser, ``{Adaptive automation for human supervision of multiple uninhabited vehicles: Effects on change detection, situation awareness, and mental workload},'' \emph{Military Psychology}, vol.~21, no.~2, pp. 270--297, 2009.

\bibitem{demir2017team}
M.~Demir, N.~J. McNeese, and N.~J. Cooke, ``Team situation awareness within the context of human-autonomy teaming,'' \emph{Cognitive Systems Research}, vol.~46, pp. 3--12, 2017.

\bibitem{wickens2017mental}
C.~D. Wickens, ``Mental workload: assessment, prediction and consequences,'' in \emph{Human Mental Workload: Models and Applications: First International Symposium, H-WORKLOAD 2017, Dublin, Ireland, June 28-30, 2017, Revised Selected Papers 1}.\hskip 1em plus 0.5em minus 0.4em\relax Springer, 2017, pp. 18--29.

\bibitem{endsley2000situation}
M.~R. Endsley and D.~J. Garland, \emph{Situation awareness analysis and measurement}.\hskip 1em plus 0.5em minus 0.4em\relax CRC press, 2000.

\bibitem{arrieta2020explainable}
A.~B. Arrieta, N.~D{\'\i}az-Rodr{\'\i}guez, J.~Del~Ser, A.~Bennetot, S.~Tabik \emph{et~al.}, ``{Explainable Artificial Intelligence (XAI): Concepts, taxonomies, opportunities and challenges toward responsible AI},'' \emph{Information fusion}, vol.~58, pp. 82--115, 2020.

\bibitem{preece2019explainable}
A.~Preece, D.~Braines, F.~Cerutti, and T.~Pham, ``{Explainable AI for intelligence augmentation in multi-domain operations},'' \emph{arXiv preprint arXiv:1910.07563}, 2019.

\bibitem{bengio2012practical}
Y.~Bengio, ``Practical recommendations for gradient-based training of deep architectures,'' in \emph{Neural networks: Tricks of the trade: Second edition}.\hskip 1em plus 0.5em minus 0.4em\relax Springer, 2012, pp. 437--478.

\bibitem{goodman2017european}
B.~Goodman and S.~Flaxman, ``{European Union regulations on algorithmic decision-making and a “right to explanation”},'' \emph{AI magazine}, vol.~38, no.~3, pp. 50--57, 2017.

\bibitem{parasuraman2010complacency}
R.~Parasuraman and D.~H. Manzey, ``{Complacency and bias in human use of automation: An attentional integration},'' \emph{Human factors}, vol.~52, no.~3, pp. 381--410, 2010.

\bibitem{jian2000foundations}
J.-Y. Jian, A.~M. Bisantz, and C.~G. Drury, ``Foundations for an empirically determined scale of trust in automated systems,'' \emph{International journal of cognitive ergonomics}, vol.~4, no.~1, pp. 53--71, 2000.

\bibitem{chung2017offensive}
T.~Chung, ``Offensive swarm-enabled tactics (offset),'' \emph{DARPA Tactical Technology Office Proposers Day, Arlington, VA, Accessed January}, vol.~30, 2017.

\bibitem{scribner2007development}
D.~R. Scribner and P.~H. Wiley, ``{The development of a virtual McKenna Military Operations in Urban Terrain (MOUT) site for command, control, communication, computing, intelligence, surveillance, and reconnaissance (C4ISR) studies},'' 2007.

\bibitem{farahani2021brief}
A.~Farahani, S.~Voghoei, K.~Rasheed, and H.~R. Arabnia, ``A brief review of domain adaptation,'' \emph{Advances in data science and information engineering: proceedings from ICDATA 2020 and IKE 2020}, pp. 877--894, 2021.

\bibitem{anderson2008asimov}
S.~L. Anderson, ``Asimov’s “three laws of robotics” and machine metaethics,'' \emph{Ai \& Society}, vol.~22, pp. 477--493, 2008.

\bibitem{murphy2009beyond}
R.~Murphy and D.~D. Woods, ``{Beyond Asimov: The three laws of responsible robotics},'' \emph{IEEE intelligent systems}, vol.~24, no.~4, pp. 14--20, 2009.

\bibitem{anderson2011machine}
M.~Anderson and S.~L. Anderson, \emph{Machine ethics}.\hskip 1em plus 0.5em minus 0.4em\relax Cambridge University Press, 2011.

\end{thebibliography}
\bibliographystyle{IEEEtran} 

\begin{IEEEbiography}[{\includegraphics[width=1in,height=1.25in,clip,keepaspectratio]{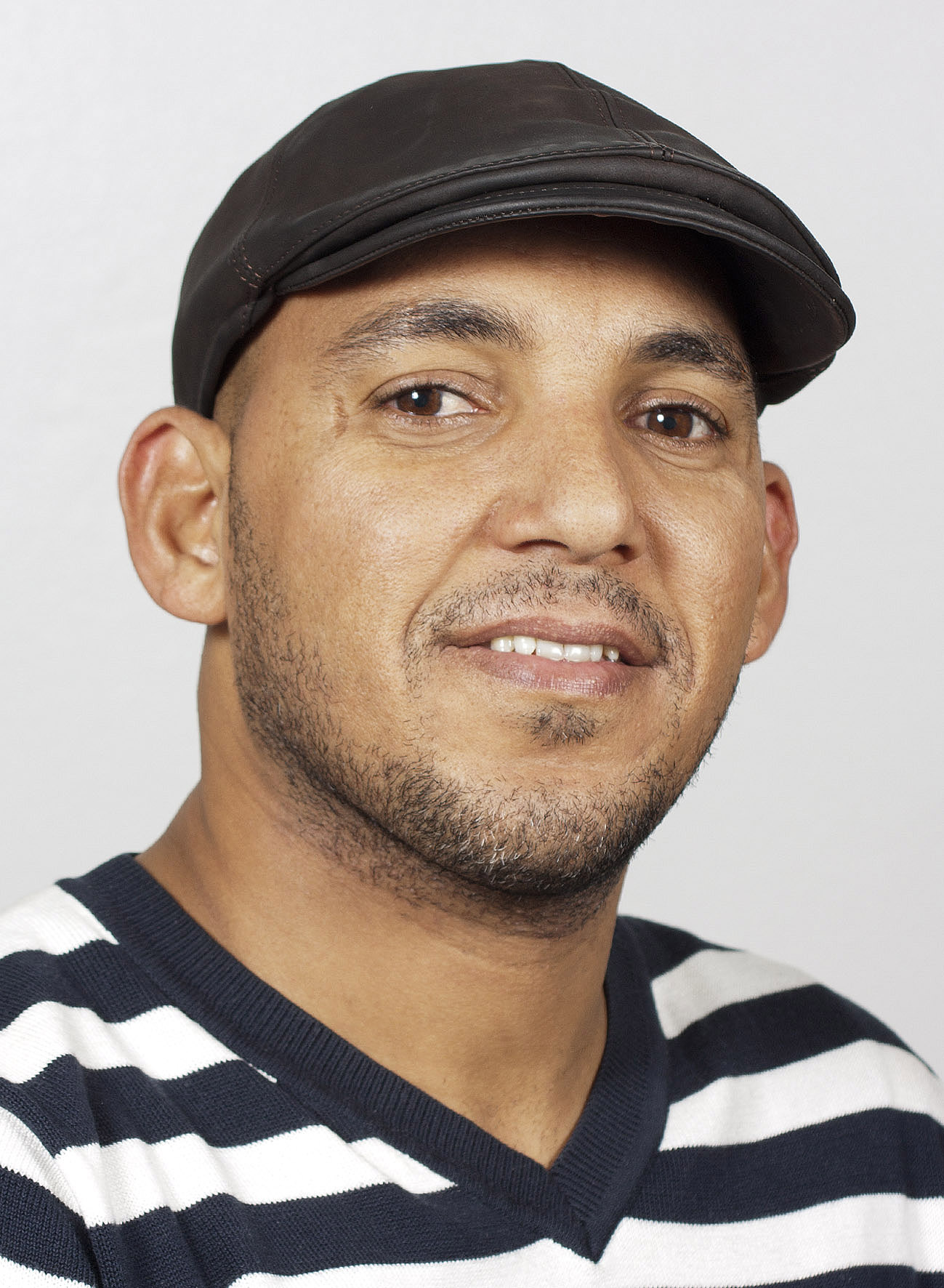}}]{Desta Haileselassie Hagos} (Member, IEEE) received a Ph.D. degree in Computer Science from the University of Oslo, Faculty of Mathematics and Natural Sciences, Norway, in April 2020. Currently, he is a Postdoctoral Research Fellow at the United States Department of Defense (DoD) Center of Excellence in Artificial Intelligence and Machine Learning (CoE-AIML), College of Engineering and Architecture (CEA), Department of Electrical Engineering and Computer Science at Howard University, Washington DC, USA. Previously, he was a Postdoctoral Research Fellow at the Division of Software and Computer Systems (SCS), Department of Computer Science, School of Electrical Engineering and Computer Science (EECS), KTH Royal Institute of Technology, Stockholm, Sweden, working on the H2020-EU project, ExtremeEarth: From Copernicus Big Data to Extreme Earth Analytics. He received his B.Sc. degree in Computer Science from Mekelle University, Department of Computer Science, Mekelle, Tigray, in 2008. He obtained his M.Sc. degree in Computer Science and Engineering specializing in Mobile Systems from Lule\aa ~University of Technology, Department of Computer Science Electrical and Space Engineering, Sweden, in June 2012. His current research interests are in the areas of Machine Learning, Deep Learning, and Artificial Intelligence.\end{IEEEbiography}

\begin{IEEEbiography}[{\includegraphics[width=1in,height=1.25in,clip,keepaspectratio]{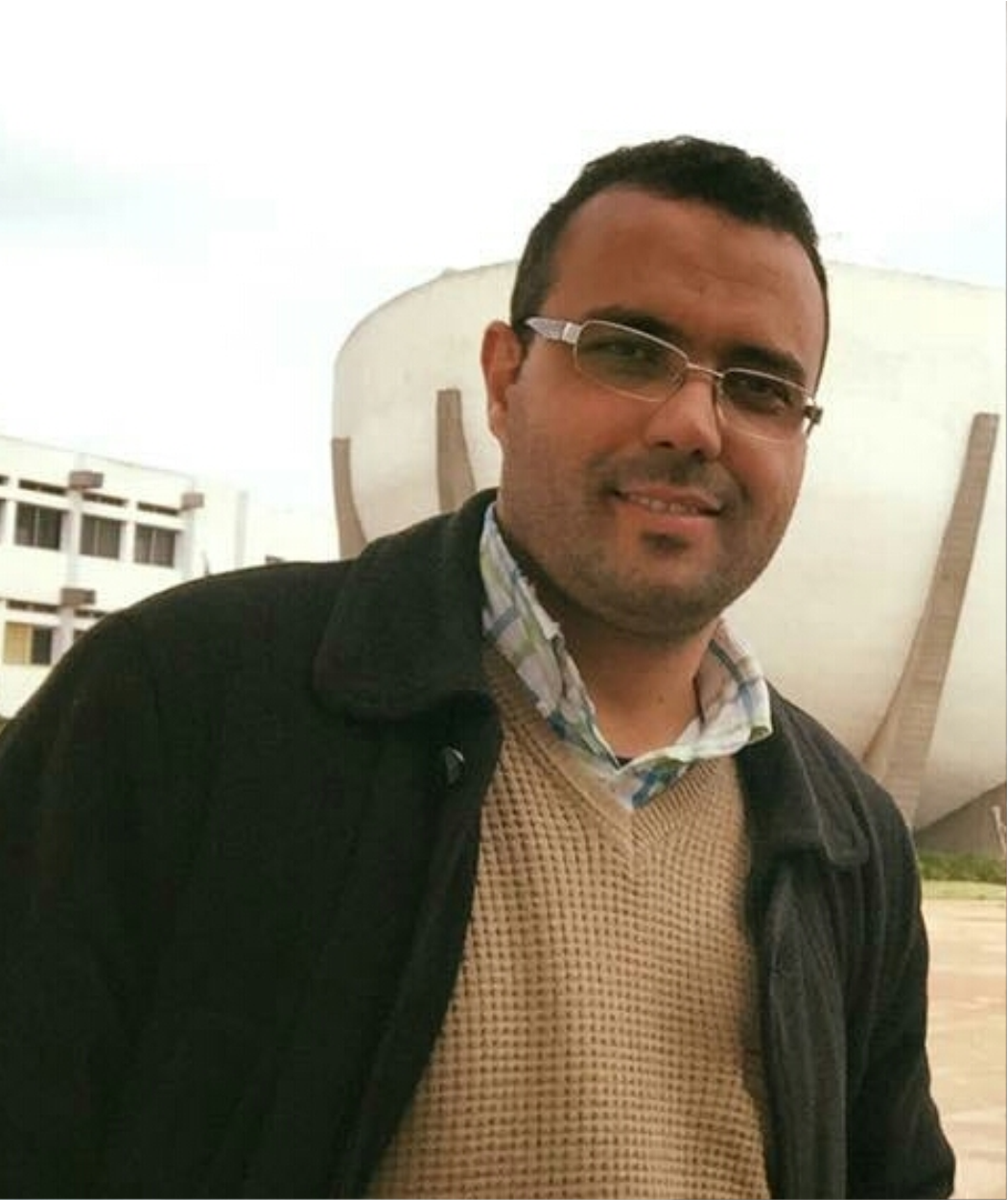}}]{Hassan El Alami} (Member, IEEE) received a Ph.D. in computer science and telecommunications from the National Institute of Posts and Telecommunications (INPT) in Rabat, Morocco, in 2019. Currently, he is a Postdoctoral Research Fellow at the DoD Center of Excellence in Artificial Intelligence and Machine Learning (CoE-AIML) at Howard University’s College of Engineering and Architecture (CEA), Department of Electrical Engineering and Computer Science in Washington, D.C., USA. Previously, he was a Postdoctoral Research Fellow at the Artificial Intelligence Research Initiative at the University of North Dakota’s College of Engineering \& Mines. His current research interests include artificial intelligence, cybersecurity, autonomous systems, the metaverse, and the Internet of Things.
\end{IEEEbiography}

\begin{IEEEbiography}[{\includegraphics[width=1in,height=1.25in,clip,keepaspectratio]{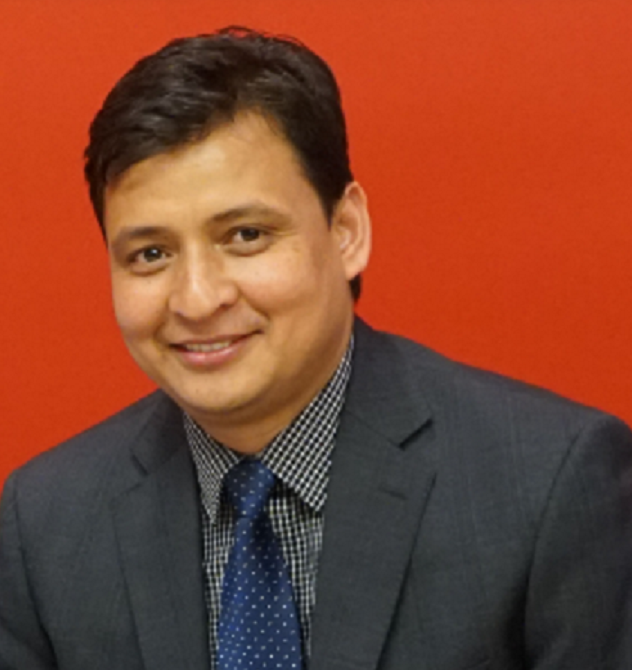}}]{Danda B. Rawat} (Senior Member, IEEE) is the Associate Dean for Research \& Graduate Studies, a Full Professor in the Department of Electrical Engineering \& Computer Science (EECS), Founding Director of the Howard University Data Science \& Cybersecurity Center, Founding Director of the DoD Center of Excellence in Artificial Intelligence \& Machine Learning (CoE-AIML), Director of Cyber-security and Wireless Networking Innovations (CWiNs) Research Lab at Howard University, Washington, DC, USA. Dr. Rawat is engaged in research and teaching in the areas of cybersecurity, machine learning, big data analytics, and wireless networking for emerging networked systems including cyber-physical systems (eHealth, energy, transportation), Internet-of-Things, multi-domain operations, smart cities, software-defined systems, and vehicular networks.Dr. Danda B. Rawat successfully led and established the Research Institute for Tactical Autonomy (RITA), the 15th  University Affiliated Research Center (UARC) of the US Department of Defense  as the PI/Founding Executive Director  at Howard University, Washington, DC, USA. Dr. Rawat is engaged in research and teaching in the areas of cybersecurity, machine learning, big data analytics and wireless networking for emerging networked systems including cyber-physical systems (eHealth, energy, transportation), Internet-of-Things, multi domain operations, smart cities, software defined systems and vehicular networks. Dr. Rawat has secured over \$110 million as a PI and over \$18 million as a Co-PI in research funding from the US National Science Foundation (NSF), US Department of Homeland Security (DHS), US National Security Agency (NSA), US Department of Energy, National Nuclear Security Administration (NNSA), National Institute of Health (NIH), US Department of Defense (DoD) and DoD Research Labs, Industry (Microsoft, Intel, VMware, PayPal, Mastercard, Meta, BAE, Raytheon etc.) and private Foundations. Dr. Rawat is the recipient of the US NSF CAREER Award, the US Department of Homeland Security (DHS) Scientific Leadership Award, Presidents’ Medal of Achievement Award (2023) at Howard University, Provost's Distinguished Service Award 2021,  the US Air Force Research Laboratory (AFRL) Summer Faculty Visiting Fellowship 2017, Outstanding Research Faculty Award (Award for Excellence in Scholarly Activity)and several Best Paper Awards. He has been serving as an Editor/Guest Editor for over 100 international journals including the Associate Editor of IEEE Transactions on Information Forensics \& Security,  Associate Editor of Transactions on Cognitive Communications and Networking, Associate Editor of IEEE Transactions of Service Computing, Editor of IEEE Internet of Things Journal, Editor of IEEE Communications Letters, Associate Editor of IEEE Transactions of Network Science and Engineering and Technical Editors of IEEE Network. He has been in Organizing Committees for several IEEE flagship conferences such as IEEE INFOCOM, IEEE CNS, IEEE ICC, IEEE GLOBECOM and so on. He served as a technical program committee (TPC) member for several international conferences including IEEE INFOCOM, IEEE GLOBECOM, IEEE CCNC, IEEE GreenCom, IEEE ICC, IEEE WCNC and IEEE VTC conferences. He served as a Vice Chair of the Executive Committee of the IEEE Savannah Section from 2013 to 2017. Dr. Rawat received the Ph.D. degree from Old Dominion University, Norfolk, Virginia in December 2010. Dr. Rawat is a Senior Member of IEEE and a Lifetime Professional Senior Member of ACM, a Lifetime Member of Association for the Advancement of Artificial Intelligence (AAAI), a lifetime member of SPIE, a member of ASEE and AAAS, and a Fellow of the Institution of Engineering and Technology (IET). He is an ACM Distinguished Speaker and an  IEEE Distinguished Lecturer (FNTC and VTS). 
\end{IEEEbiography}

\end{document}